\documentclass[aps,showkeys,floatfix,nofootinbib,twocolumn,amsfonts,amsmath,longbibliography,prb]{revtex4-2}

\usepackage{CJK}
\usepackage{pgf}
\usepackage{physics}
\usepackage[colorlinks=true,hyperfootnotes=true,breaklinks=true,citecolor=blue,linkcolor=blue,urlcolor=blue]{hyperref}

\usepackage[defaultcolor=red,final]{changes} 

\begin{document}

        \title{On the dual character of Zn impurity in SnTe: Tuning thermoelectric and\\ topological properties}
        \author{Kacper Pryga}
        \email{pryga@agh.edu.pl}
        \author{Bartlomiej Wiendlocha}
        \email{wiendlocha@fis.agh.edu.pl}
        \affiliation{AGH University of Krakow, Faculty of Physics and Applied Computer Science, Aleja Mickiewicza 30, 30-059 Krakow, Poland}

        \begin{abstract}
We present a first-principles study of the electronic structure and thermoelectric properties of Zn-doped SnTe using the Korringa-Kohn-Rostoker method within the coherent potential approximation, complemented by pseudopotential calculations. 
SnTe is a lead-free analogue of PbTe and a candidate thermoelectric material in which Zn doping has been experimentally reported to enhance the performance of $p$-type samples.
We show that Zn introduces a resonant-like impurity state, located within the conduction band, which evolves strongly depending on the Zn concentration. This feature leads to a significant enhancement of the thermopower in $n$-type SnTe. In the valence band, Zn doping induces L–$\Sigma$ band convergence, also resulting in an increased $p$-type Seebeck coefficient over a broad concentration range and delaying the onset of the bipolar effect.
We further demonstrate that the Zn-induced band-structure modifications drive a transition from an inverted to a trivial band ordering, indicating a controllable topological phase transition. Our results clarify the microscopic role of Zn in SnTe and identify doping as a mechanism for simultaneously tuning thermoelectric and topological properties.
        \end{abstract}
        
        \date{\today}

        \maketitle

    
\section{Introduction}

Current energy concerns spark the global search for an efficient and environmentally friendly alternative to conventional energy sources. Thermoelectric (TE) materials capable of direct conversion of heat into electricity without any emissions appear to be one of the most viable solutions to these challenges. The absence of moving parts, resulting in low maintenance and high reliability, makes them useful for powering spacecraft or a promising candidate for waste heat recovery applications \cite{Goldsmid2010IntroductionThermoelectricity}. Unfortunately, in comparison to conventional solutions, the low efficiency of TE materials restricts their range of applications.

TE efficiency depends on the dimensionless figure of merit $ZT=\frac{S^2\sigma}{\kappa_l+\kappa_e}T$, where $S$, $\sigma$, $T$, $\kappa_l$, $\kappa_e$ are thermopower, electrical conductivity, temperature, lattice and electronic thermal conductivity, respectively. Among these parameters, only $\kappa_l$ remains mostly independent, while $S$, $\sigma$, and $\kappa_e$ are strongly coupled through electronic structure \cite{Goldsmid2010IntroductionThermoelectricity}. 
Thus, to obtain a high $ZT$, the reduction of $\kappa_l$ plays a major role and can be achieved, for example, by introducing point defects, dislocations, or by using nanostructuring \cite{Li2019SnTe-BasedThermoelectrics}. Unfortunately, this approach is limited by the amorphous limit, meaning that increasing the efficiency of TE materials must be achieved by directly altering the carrier concentration and finding the optimal ratio of the above-mentioned, intertwined parameters. This often is a challenging process, and while doping for many years has been the main procedure of modifying $n$ and through it the $S$, currently various band engineering strategies are often used, e.g. band convergence, nesting, increasing band degeneracy and introducing resonant states \cite{Li2019SnTe-BasedThermoelectrics}.

Resonant levels (RLs) have been known in metals for more than 60 years (as ''virtual bound states'' \cite{Friedel1956ONSOLUTIONS}), yet only recently have RLs been rediscovered in semiconducting thermoelectrics~\cite{Heremans2008EnhancementStates}, as their introduction can substantially enhance the thermopower of TE materials in a high carrier concentration range. 
RLs arise from the introduction of impurity atoms with energy levels that retain some of the atomic-like characteristics of having a small energy width yet being located within the band of the host semiconductor and having a necessary degree of hybridization. As a result, RL distorts the electronic structure of the host material \cite{Heremans2012ResonantSemiconductors,Heremans2008EnhancementStates}. 
Some examples of successful thermopower enhancements by RL have been reported for various dopants in highly efficient TE materials, namely Tl in PbTe  \cite{Heremans2008EnhancementStates,Heremans2012ResonantSemiconductors,Parashchuk2021HighSoftening}, 
Sn in $\beta$-$\mathrm{As_2Te_3}$ \cite{Wiendlocha2018Ansub3RLinsnas2te3}, Sn in $\mathrm{Bi_2Te_3}$ \cite{Jaworski2009ResonantPower,Wiendlocha2016ResonantTetradymites} or In in  GeTe \cite{Wu2017ResonantGeTe} and SnTe~\cite{Zhang2013HighSnTe,Misra2020BandLevel,Misra2022Influence/math}.

The enhancement in thermopower by RLs is accompanied by several interwoven effects: an increase in the density of states (DOS) around the energy of the RL, an increase of the DOS effective mass, and the presence of resonant scattering of carriers as a result of strong band smearing~\cite{Wiendlocha2013FermiCalculations,Wiendlocha2018Thermopower,Wiendlocha2021ResidualSemiconductors}.
The plot of the DOS has been a standard RL identification technique from a theoretical point of view because of the formation of a sharp peak in the impurity states.
However, not every peak in DOS is a resonant level, and not every
resonant level will positively affect the thermopower and power factor of TE materials. 
For a better characterization of an impurity state whether it can be considered resonant or not, beyond the DOS plot in theoretical calculations, the Bloch spectral density functions~\cite{Faulkner1980CalculatingApproximation, Ebert2011CalculatingApplications} are probably the most useful tool. They allow one to investigate whether the impurity states tend to form an impurity band or strongly smear the host band structure. Moreover, the energy-dependence of the spectral function at a selected $\vb{k}$-point for a resonant states loses the Lorentzian shape~\cite{Wiendlocha2013FermiCalculations,Wiendlocha2021ResidualSemiconductors}, which gives strong arguments on their non-rigid-band-like character.
Analysis of the width of the spectral functions also allows one to estimate the electronic lifetime, which becomes short in the presence of resonant scattering~\cite{Gyorffy1979FirstAlloys}.
This observation allows for an experimental investigation of the resonant character of an impurity, independent of its effect on the thermopower of the material by measuring the resistivity and mobility at low-temperatures~\cite{Wiendlocha2021ResidualSemiconductors}.

A very informative example of a resonant state that does not increase the thermopower and shows the benefit of using the spectral function technique is PbTe doped with titanium~\cite{pbte-ti-konig, Heremans2012ResonantSemiconductors}. Here, the 3d orbitals of Ti electrons form a resonant level in the conduction band of PbTe, but no thermopower enhancement is experimentally observed. 
This was explained~\cite{Wiendlocha2014LocalizationPbTe} as an effect of the formation of impurity bands detected by the calculated Bloch spectral functions, since the 3d states of Ti formed dispersionless impurity band-like states in $\vb{k}$-space, not hybridized with the host PbTe valence band. As a result, the localized impurity band had no contribution to the total thermopower of the material because the thermopower of such a two-band system is a weighted average of the thermopowers of each individual band, with conductivities as weights. The contribution of a localized impurity band, which has a small conductivity, is then averaged, and only the original conduction band of PbTe contributed to the thermopower of the material.
This showed that for RL to be effective in enhancing the thermopower, an impurity band should not be formed, and the resonant states should be hybridized with the electronic states of the host material.

Careful optimization is also needed, 
as too strong hybridization, realized e.g. by too large impurity concentration or chemical pressure~\cite{jaworski-ees,Heremans2012ResonantSemiconductors} wipes out the resonant nature of the impurity decreasing excess in thermopower.
Balance must also be provided between the RL-induced band smearing effect and band convergence when trying to use both techniques to further enhance the TE performance, as too strong band smearing diminishes the positive effect of band convergence~\cite{snte-misra2024}.

The current work is devoted to the analysis of the character of a Zn dopant in SnTe, a member of the chalcogenide family of thermoelectric materials, a more environmentally-friendly alternative for PbTe~\cite{Li2019SnTe-BasedThermoelectrics,Moshwan2019RealizingEngineering,Chen2020RoutesThermoelectrics,Banik2014LeadfreeSystem,Freer2022}. 
Although it has the same crystal structure and a very similar band structure as PbTe, pristine SnTe has a much lower $ZT$ of 0.5 at 900 K \cite{Zhou2014OptimizationBand}.
This comes mainly from three effects: a
large number of Sn vacancies leading to a high carrier concentration, a higher thermal conductivity $\kappa_l\approx 2 - 3$ $\mathrm{Wm^{-1}K^{-1}}$ \cite{Zhou2014OptimizationBand,Pei2011HighPbTe} and a large energy offset $\approx0.3$ eV between the maxima of the valence band (VB) at L and $\Sigma$ points \cite{Rogers1968ValenceSnTe}. 
The consequence of the large offset is that the high degeneracy of the $\Sigma$ band ($N_V=12$) is not utilized in the carrier transport. For SnTe, in order to counter those effects and to be competitive with its counterpart, the simultaneous employing of multiple band engineering methods is often required.

The resonant impurity of In was successfully used to improve the thermoelectric performance of SnTe~\cite{Zhang2013HighSnTe,Tan2016Element-selectiveTl,Misra2020BandLevel}. Introducing 0.25 atomic \% of In in SnTe has increased $ZT$ to 1.1 at 873 K \cite{Zhang2013HighSnTe}. Recent years have brought multiple successful attempts to achieve higher $ZT$ by co-doping In with various other elements \cite{Wang2017EnhancementSoftening,Wang2017ManipulatingCodoping,Banik2014LeadfreeSystem,Wang2018OptimizationAlloys,Wang2019EnhancingProperties,Banik2016HighConvergence,Li2018EnhancementCoDoping,Doi2019BandSnTe,Tan2015HighScattering,Guo2019SimultaneousSnTe,Guo2019SynergisticSnTe,Hussain2020RealizingCu2Te,Lu2020ThermoelectricMethod,Moshwan2019RealizingEngineering,Srinivasan2018ThermoelectricSintering,Tan2015CodopingConvergence,Zhang2022Super-structuredMaterials,Wang2019ThermoelectricLimit,Pang2022RealizingDistortion,Wang2020HierarchicalThermoelectrics,Zhang2018EnhancedCo-doping,Zhou2016ThermoelectricLevels,Heo2023,Wang2025}, and the co-doped Mn-In system probably showed the highest $ZT$ value of 1.30 at 850 K~\cite{snte-misra2024}.

Moreover, several new resonant dopants were suggested.
Examples (both single and co-doped) of promising RL-candidates are Bi \cite{Kihoi2021OptimizedTuning,Shenoy2021ImprovingCo-dopant,Shenoy2019ElectronicTelluride,Yang2021RealizingSnTe,Zhang2021EnhancedDefects,Hong2023,Ganesan2025,Xia2023} as well as Zn, which is the main focus of this study. 
The first reports of Zn being the resonant dopant appeared in 2019 \cite{Bhat2019Zn:Thermoelectrics}. Experimental Zn-doped samples had room temperature carrier concentration in a range of $1.6 - 2.05\times10^{20}$ cm$^{-3}$ (for 2\% to 8\% Zn atomic concentration). The highest thermopower of $\sim127$ $\mu$VK$^{-1}$ at 300 K was obtained for $\mathrm{Sn_{0.96}Zn_{0.08}Te}$.

Previous studies on SnTe:Zn focused mainly on other effects of Zn doping, such as lowering thermal conductivity, widening the band gap, and band convergence \cite{Aminzare2019EffectSnTe,Chen2018BandZn-doping,Dong2016First-principlesDoped-SnTe,Wang2024}. Furthermore, several studies were conducted for co-doped variants of SnTe with Zn~\cite{Bhat2020SnTePerformance,Zhang2021EnhancedDefects,Shenoy2020BiiZT/i,Shenoy2022ACo-dopants,Kihoi2020NanostructuringCo-doping,Wang2025,Bugalia2024}. 
Despite this, the electronic structure of \replaced[comment=R1Q1]{single-doped SnTe with Zn}{SnTe:Zn} has not been studied in detail 
\added[]{as in most of the works where DFT calculations were present, only a single supercell was studied that corresponded to a  relatively large concentration (at least ~3.1\% Zn) \cite{Zhang2021EnhancedDefects,Bhat2019Zn:Thermoelectrics,Dong2016First-principlesDoped-SnTe,Shenoy2020BiiZT/i}. Among those works
Zhang et al. \cite{Zhang2021EnhancedDefects} performed calculations for a fully-relaxed 2$\times$2$\times$2 supercell with composition Sn$_{31}$ZnTe$_{32}$ corresponding to 3.1\% of Zn concentration. Calculations on a similar structure were performed in work \cite{Bhat2019Zn:Thermoelectrics} together with a smaller 2$\times$2$\times$1 supercell Sn$_{15}$ZnTe$_{16}$ (6.3\% Zn equivalent). Similar results were presented in work \cite{Shenoy2020BiiZT/i}, where the Authors performed calculation of fully-relaxed $\sqrt{2}$$\times$$\sqrt{2}$$\times$2 supercell Sn$_{15}$ZnTe$_{16}$. Finally, Dong et al. \cite{Dong2016First-principlesDoped-SnTe} and Wang et al. \cite{Wang2024} have shown calculations for  3$\times$3$\times$3 Rock-Salt supercell with 54 atoms which corresponds to the structure considered in this study. It is worth pointing out that in Refs. \cite{Dong2016First-principlesDoped-SnTe,Wang2024} the supercells were based on a primitive face-centered cell, while in the remaining works, the simple cubic structures were used. All of the available in the literature calculations on SnTe:Zn were performed using the pseudopotential method and by utilizing the GGA-PBE exchange-correlation potential. 
A more detailed discussion regarding the existing DFT studies and the discrepancies present will be provided later in a section devoted to the electronic structure}.
The question of whether Zn forms a resonant level in SnTe and whether it is responsible for the improved $p$-type thermoelectric performance of this material is addressed here.

In this paper, we present a detailed analysis of the electronic structure of SnTe:Zn, obtained via \textit{ab-initio} calculation, focusing  on the formation of RL, as well as the modification and evolution of the band structure under the influence of Zn doping.
We show that Zn creates a resonant level in the conduction band, and at the same time has a strong influence on the valence band of SnTe, especially in the L$-\Sigma$ direction. This, together with a decrease in the offset between valence band maxima, and the increased carrier mass resulting from substantial band flattening, explains the enhanced thermopower, as seen in the experiment \cite{Bhat2019Zn:Thermoelectrics}. These findings are supported by calculations of the Seebeck coefficient based on the Kubo-Greenwood formalism for both $p$- and $n$-type SnTe:Zn, which predicts a vastly improved thermoelectric performance, allowing further improvement of SnTe-based thermoelectric materials.
We further demonstrate that the Zn-induced band-structure modifications drive a transition from an inverted band ordering to a trivial band ordering, indicating a controllable topological phase transition, and making Zn-doped SnTe an interesting system for studying topological properties.
       

\section{Computational methods}

\begin{figure*}
    \includegraphics[width=\textwidth]{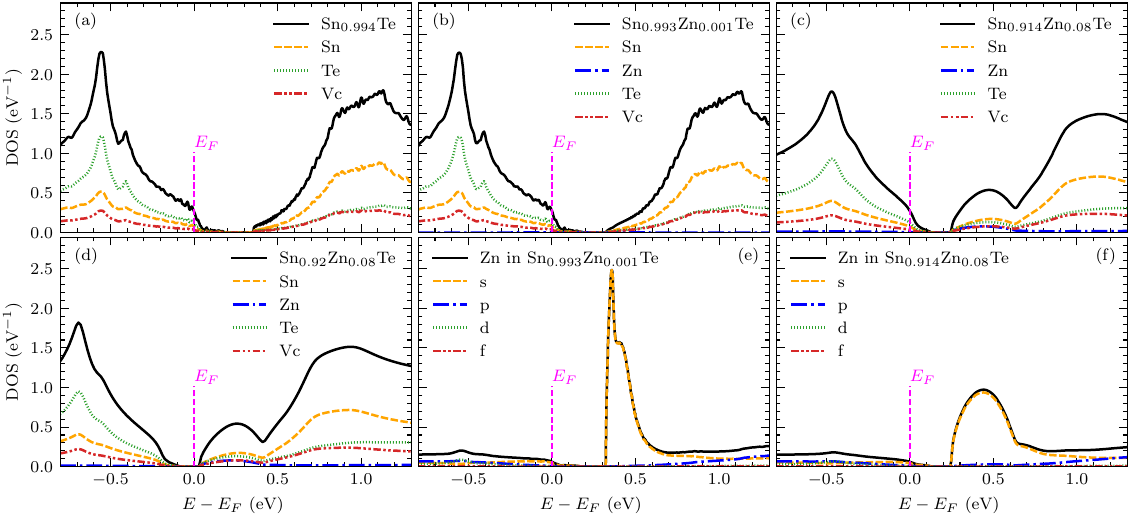}
    \caption{Calculated density of states (DOS) near $E_F$ for SnTe with 0.6\% Sn vacancies: (a) without doping; (b) doped with 0.1\% Zn, with partial density of states (PDOS) in panel (e); (c) doped with 8\% Zn, with PDOS in panel~(f). In panel (d) SnTe with 8\% Zn without vacancies is shown. Partial DOS (panels d and e) is weighted per impurity atom (i.e. does not include concentration).}
    \label{fig:001a_dos_comp}  
\end{figure*}

The electronic-structure calculations were carried out using the full-potential Korringa–Kohn–Rostoker (KKR) method in combination with the coherent potential approximation (CPA) to model the effects of chemical disorder, as implemented in the Munich {\sc spr-kkr} code \cite{Ebert2011CalculatingApplications,Ebert2019TheMunich}.
Computations were performed in a fully relativistic manner, that is, the spin-orbit interaction was included. The crystal potential was constructed using the local density approximation (LDA) with Vosko, Wilk, and Nussair parameterization \cite{Vosko1980AccurateAnalysis}.
The position of the Fermi level was obtained using the Lloyd formula \cite{Ebert2011CalculatingApplications} and the angular momentum cutoff was set to $l_{max}=3$. The threshold for the convergence of the self-consistent cycle was set at $5\times10^{-4}$ Ry for the total energy. For a self-consistent cycle, regular mesh of about 8000 $\vb{k}$-points (within the irreducible part of Brillouin zone) was used, and $1.5-3\times10^5$ $\vb{k}$-points for the densities of states (DOS) and the Bloch spectral density functions (BSFs). 
The experimental crystal structure \cite{Bhat2019Zn:Thermoelectrics} (NaCl type, space group 225) was used with a lattice parameter of $a=6.315 - 6.327$ \AA\ linearly depending on the Zn concentration.

BSFs calculations were performed to further study the modifications of the electronic structure arising from doping. BSFs in disordered systems serve as more generalized dispersion relations \cite{Ebert2011CalculatingApplications,Ebert1997RelativisticAlloys,Faulkner1980CalculatingApproximation}. 
For a regular ordered crystal, BSF at a given $\vb{k}$ is a Dirac delta function of energy $\delta(E-E_{\vb{k}})$, showing the position of the band. 
Scattering of electrons in a system with impurities leads to a perturbation of the electronic structure, and bands become smeared. For nonresonant impurities and when the impurity concentration is not high, the BSF at a single $\vb{k}$-point takes the form of the Lorentz function~\cite{Gyorffy1979FirstAlloys,Gordon1981OnAlloys,PhysRevB.31.3260,PhysRevB.29.4217},
\begin{equation}
    A^B(\vb{k},E)=L(E)=\dfrac{1}{\pi} \dfrac{\frac{1}{2}\Delta}{(E-E_0)^2+(\frac{1}{2}\Delta)^2},
\end{equation}
where the peak marks the center of the smeared band and the width $\Delta$ corresponds to the lifetime of the electronic state: 
\begin{equation}
\tau=\hbar/\Delta.
\end{equation}

As mentioned above, BSFs have so far been successfully applied multiple times while studying the character of resonant impurities in TE materials  \cite{Kim2014Electronic0.1,Wiendlocha2018Thermopower,Wiendlocha2013FermiCalculations,Wiendlocha2014LocalizationPbTe,Misra2020BandLevel,Parashchuk2024,Pryga2025}. An advantage of this technique is the possibility to better characterize the resonant or non-resonant character of the dopant by analysis of the BSF shape. Furthermore, in a case of resonant impurities, BSFs allow us to study the tendencies towards localization of the resonant states. This is crucial from the perspective of TE properties, as localization hampers the improvement of the Seebeck coefficient  \cite{Heremans2012ResonantSemiconductors}.

The calculation of transport properties was performed within the Kubo-Grenwood formalism \cite{Kubo1957,Greenwood1958,PhysRevB.31.3260,Swihart1986} in combination with the KKR-CPA method~\cite{Ebert2011CalculatingApplications,Ebert2019TheMunich,Kdderitzsch2011,Popescu2017}. 
Within this approach, it is possible to obtain the absolute values of conductivity and thermopower while accounting for the chemical disorder--induced electron scattering effects. For the calculation of the energy-dependent transport function $\sigma(E)$ a mesh of $5\times 10^{4}-10^{7}$  $\vb{k}$-points was used. An energy span of $\pm0.7$ eV around $E_F$ was considered, with a small energy step of 0.2 mRy. In the calculation of $\sigma(E)$ atomic-sphere approximation (ASA) was used for the potential.

To confirm the observed modifications of the band structure and check for the possible effect of relaxation of the local crystal structure around impurity atoms, calculations were also performed using the supercell method. For this purpose, the pseudopotential method, as implemented in the Vienna \textit{ab initio} simulation package ({\sc VASP}) \cite{Kresse1993iAbMetals,Kresse1994iAbGermanium,Kresse1996EfficiencySet,Kresse1996EfficientSet,Kresse1999FromMethod}, was used, with the Projector Augmented Wave (PAW) pseudopotentials \cite{Blochl1994ProjectorMethod} and the generalized gradient approximation with the Perdew-Burke-Ernzerhof (PBE) exchange-correlation functional \cite{Perdew1996GeneralizedSimple}.   
A 3$\times$3$\times$3 supercell (54 atoms) was constructed from the primitive cell of SnTe. 
A single Sn atom was substituted with Zn, simulating the $\mathrm{Sn_{0.963}Zn_{0.037}Te}$ structure. 
The cutoff energy for the plane wave base was set at 320 eV, and for the self-consistent cycle, the energy convergence threshold was set at $10^{-10}$ eV.
We used a Monkhorst-Pack grid of 5$\times$5$\times$5 $\vb{k}$-points for the self-consistent cycle and 8$\times$8$\times$8 for calculations of the DOS and band structure. 
For the relaxation of atomic positions and cell shape, strict convergence limits on forces of 1 meV/\AA\  were used.
The unfolding of the supercell band structure was performed using the {\sc VASPBandUnfolding} package \cite{Zheng2022VASPBandUnfolding}. \replaced[comment=R3Q4]{For band parity analysis the {\sc C2x} software was used \cite{Rutter2018}. Calculation of the mirror Chern number $n_M$ was performed using the {\sc Z2Pack} \cite{Gresch2017,Soluyanov2011} which is interfaced with the {\sc Quantum ESPRESSO} package \cite{Giannozzi2017,Giannozzi2009}. Here, the calculations were performed using fully-relativistic SG15 Optimized Norm-Conserving Vanderbilt pseudopotentials \cite{Hamann2013,Scherpelz2016} with PBE exchange-correlation functional \cite{Perdew1996GeneralizedSimple}. Kinetic
energy cutoffs in the plane wave expansion of wave functions and charge density were set to 70 Ry and 280 Ry, respectively with the convergence tested for cutoffs 1.5 times larger. For the self-consistent cycle, Monkhorst-Pack grid of 5$\times$5$\times$5 $\vb{k}$-points was used in the case of Sn$_{26}$ZnTe$_{27}$ and 20$\times$20$\times$20 $\vb{k}$-points for SnTe.}


\section{Results and discussion}

\subsection{Density of states}

Figure~\ref{fig:001a_dos_comp} presents the computed electronic density of states (DOS) of SnTe with and without Zn impurity. 
To take into account that the synthesized samples of SnTe always contain $p$-type Sn vacancies, 0.6\% of vacancies on the Sn site were included in each calculated variant. This amount of vacancies, which here act as two-hole acceptors, corresponds to the carrier concentration of $1.9\times 10^{20}$ cm$^{-3}$, close to that observed experimentally~\cite{Bhat2019Zn:Thermoelectrics}. \added[comment=R2Q1]{Abundance of Vc$_{\rm Sn}$ in SnTe stems from very low, negative formation energy in contrast to the Te vacancies which have formation energies above 1 eV making it difficult to form \cite{Wang2014}. Presence of Te vacancies would result in the n-type material, which was (unfortunately) not yet reported.} 
Sn vacancies, in agreement with earlier reports \cite{Misra2020BandLevel}, lead to a rigid band-like shift of the Fermi level deeper into the valence band, as seen in Fig.~\ref{fig:001a_dos_comp}(a).
Calculating DOS at low dopant concentration is often used as a first indication of the possible resonant character of an impurity, as a narrow peak in the partial density of states is formed for RL~\cite{Heremans2012ResonantSemiconductors,Misra2020BandLevel, Wiendlocha2016RecentMaterials}. With this in mind, we performed calculations of the electronic structure of SnTe doped with 0.1\% Zn on the Sn site. 
Although the total DOS, shown in Fig.~\ref{fig:001a_dos_comp}(b), is not modified due to the low impurity content, the partial density of states (PDOS) of Zn, presented in Fig.~\ref{fig:001a_dos_comp}(e), clearly shows the formation of a peak at the bottom of the conduction band (CB). 
The peak is not as narrow as for 0.1\% of Tl in PbTe~\cite{Heremans2012ResonantSemiconductors,Wiendlocha2016RecentMaterials} but indicates a possibility of RL formation in the conduction band. 

For higher concentrations of Zn, in particular in the case of \replaced[comment=R2Q2]{$\mathrm{Sn_{0.914}Zn_{0.08}Te}$}{$\mathrm{Sn_{0.914}Sn_{0.08}Te}$} [Fig. \ref{fig:001a_dos_comp}(c)], the obtained DOS reveal that increasing the concentration of impurity strongly affects the conduction band of SnTe. The DOS hump is now clearly visible in the total DOS, which arises from hybridization of Zn 4s orbitals with s and p orbitals of Sn and Te. Here, the contributions at the resonant peak result mainly from Sn and Te atoms (37\% and 29\%, respectively), and Zn makes only about 17\% of total DOS. 
This shows that the presence of Zn "catalyzes" the reconstruction of DOS due to strong hybridization with the host atoms' orbitals, proving the delocalized nature of this resonant level.

Fig.~\ref{fig:001a_dos_comp}(d) shows the important feature of the Zn impurity. When the system contains no Sn vacancy, the Fermi level is located in the gap. Zn is a divalent element, similar to Sn; therefore, the substitution of Zn$_{\rm Sn}$ has no effect on the number of carriers and the position of the Fermi level. The $p$-type character of the Sn$_{0.994-x}$Zn$_x$Te samples comes only from Sn vacancies. 

\begin{figure}
    \includegraphics[width=0.5\textwidth]{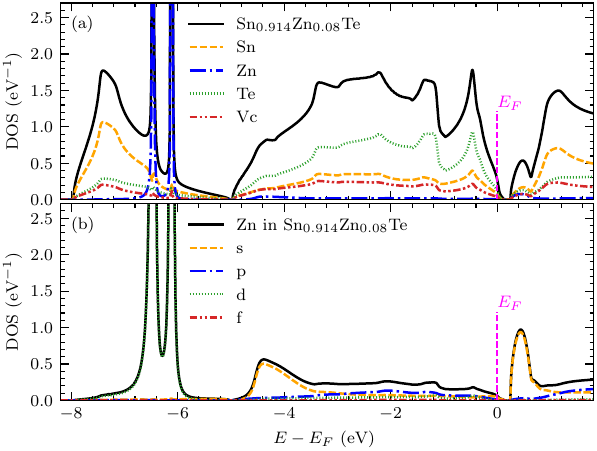}
    \caption{(a) Calculated density of states (DOS) for $\mathrm{Sn_{0.914}Zn_{0.08}Te}$ in broad range of energy  with partial DOS of Zn (panel b)  weighted per impurity atom.}
    \label{fig:001b_dos_comp}    
\end{figure}

Figures~\ref{fig:001b_dos_comp}(a) and \ref{fig:001b_dos_comp}(b) show total and partial DOS in a wider energy range. Deep within the valence band, below $-4$ eV, the second resonant-like (hyperdeep, HD) peak in the DOS of 4s states of Zn appears, similar to what we know from e.g. Tl in PbTe~\cite{Heremans2012ResonantSemiconductors,mahanti-pbte-prl,mahanti-pbte-prb}.
Below that, around \replaced[comment=R2Q2]{$-6$}{$-7$} eV, the semi-core 3d states of Zn, split due to the spin-orbit coupling, are located.

The calculated DOS of the KKR-CPA method is confirmed by the supercell calculations; 
comparison is shown in the Supplemental Material, Fig. S1~\cite{suppl}. In particular, the local structural relaxation around Zn does not remove the peak in DOS.

\begin{figure}
    \centering
    \includegraphics[width=0.5\textwidth]{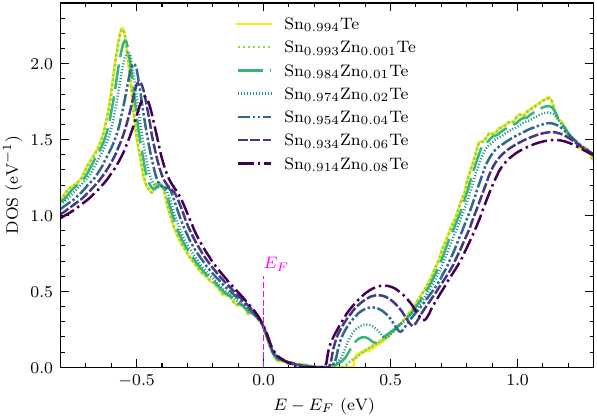}
    \caption{Evolution of total DOS near $E_F$ for Sn$_{0.994-x}$Zn$_{x}$Te with Zn doping.}
    \label{fig:002_dos_compilation}
\end{figure}

Figure~\ref{fig:002_dos_compilation} presents the total DOS near the band gap for a range of Zn concentrations, showing the evolution of the electronic structure with doping. The Sn vacancy concentration was kept constant at 0.6\%, so the Fermi level does not move.
With increasing amounts of Zn, the resonant ''hump'' in DOS grows further, while its maximum is moved deeper into the conduction band. This is possible because of the redistribution of the available electronic states, as neutral Zn does not increase the number of electrons and the number of available electronic states. As a consequence, the DOS at higher energies decreases, transferring the electronic states to the RL hump.
This shows that the electronic structure in the $\vb{k}$-space must have changed as the number of available states in the energy interval of approximately 0.3 eV above the CB edge has increased substantially. This subject is analyzed using the spectral functions technique in the following paragraph.
\added[comment=R3Q7]{Considering a steady growth of the DOS at the resonant peak with impurity content, we can additionally expect an enhancement of the electron-phonon interaction strength with Zn doping in this energy region.}

On the other hand, the DOS of the valence band shows less spectacular modifications. Also there we observe a redistribution of electronic states, as the DOS becomes steeper and larger below $E_F$, indicating the shift of bands towards $E_F$, which is an indirect indication of the band convergence effect.

\replaced[comment=R1Q1]{As far as comparison with the literature is concerned, i}{I}n \added[]{majority of the} previously reported supercell calculations for Zn-doped SnTe \cite{Bhat2019Zn:Thermoelectrics,Shenoy2020BiiZT/i,Zhang2021EnhancedDefects,Wang2024}, the band gap was closed by the Zn peak, making it impossible to conclude whether the RL is in the valence or conduction band. \added{The  RL peak present in the calculations reported in \cite{Bhat2019Zn:Thermoelectrics,Shenoy2020BiiZT/i,Wang2024} was also relatively small with respect to the rest of the band and not as noticeable as in this work (both calculated using the CPA approach or supercell - see Fig. S1 in Ref.~\cite{suppl}) or in \cite{Dong2016First-principlesDoped-SnTe}. This discrepancy in character of the DOS peak might primarily stem from the computational details, e.g. choice of smearing and density of the k-points mesh. Other possible sources of discrepancies will be discussed later.}
One theoretical work reported the formation of an impurity level at the bottom of the conduction band \cite{Dong2016First-principlesDoped-SnTe} \added[]{which manifests itself as a sharp peak in the DOS, akin to the one obtained for $\mathrm{Sn_{26}ZnTe_{27}}$ supercell (Fig. S1~\cite{suppl}), although in the mentioned work the DOS peak is slightly detached from the conduction band itself, while our calculations predict the overlap of RL peak with the conduction band}.

\replaced[]{Overall, a}{A}nalysis of DOS shows that the resonant level located in the conduction band cannot be directly responsible for the reported~\cite{Bhat2019Zn:Thermoelectrics} enhancement of thermopower in $p$-type Zn doped SnTe. RL in the conduction band could influence the $p$-type thermopower through the bipolar effect at high temperatures; thus the experimentally observed enhancement of the thermopower at room temperature does not originate directly from RL. Now we proceed to the analysis of the spectral functions, which sheds more light on the Zn behavior.  
\begin{figure*}
    \centering
    \includegraphics[width=\textwidth]{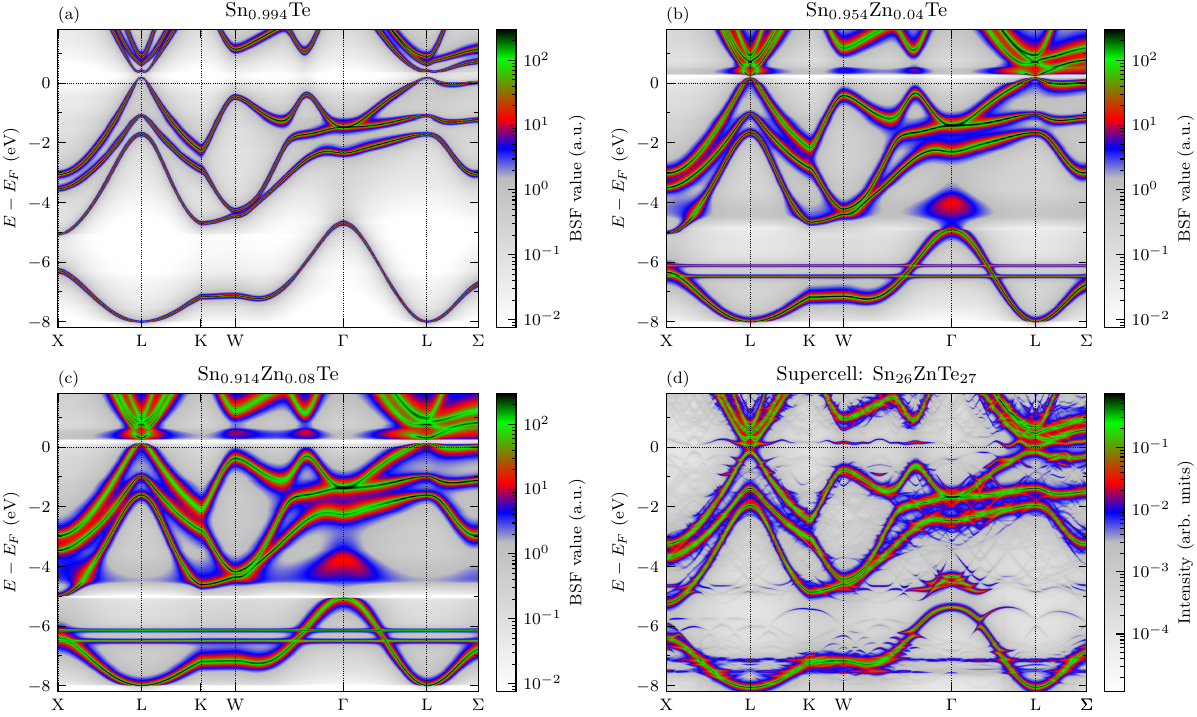}
    \caption{Two-dimensional projections of Bloch spectral functions for $\mathrm{Sn_{0.994}Te}$: (a)  without doping; (b) doped with 4\% Zn; (c) doped with 8\% Zn. 
    Panel (d) presents unfolded bands from the  $\mathrm{Sn_{26}ZnTe_{27}}$ supercell.
    Both BSFs, and unfolded bands were plotted along high symmetry directions in the Brillouin Zone. Color representing BSF is in a logarithmic scale, where black color corresponds to values higher than 300 atomic units. Similarly, intensity of the unfolded bands is also in logarithmic scale. Note substantial smearing of bands at the L and $\Sigma$ points resulting from the resonant level scattering at the edge of the conduction band. In both approaches, no signs of a localized impurity band are present \added{near the Fermi level}.}
    \label{fig:003_bsf_full}
\end{figure*}

\begin{figure*}
    \centering
    \includegraphics[width=\textwidth]{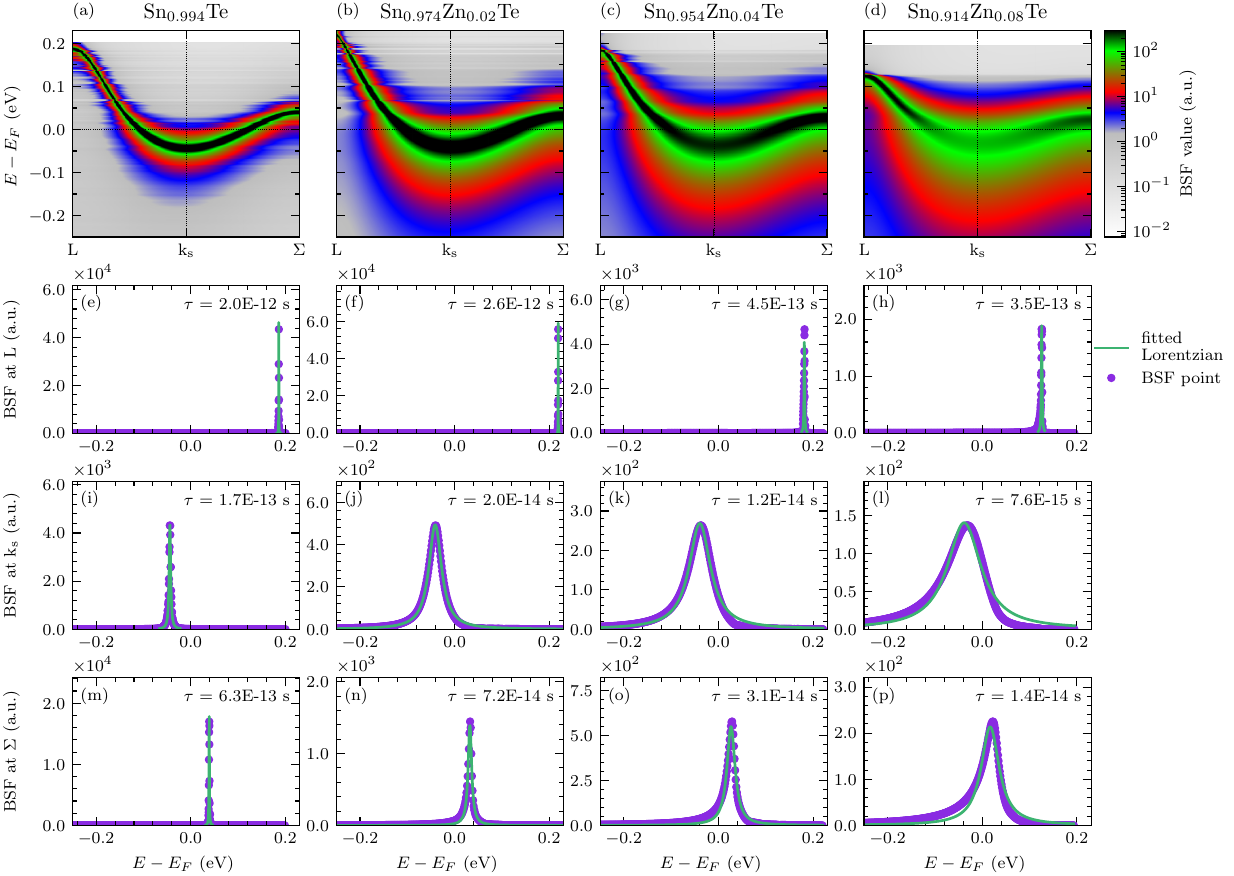}
    \caption{Calculated Bloch spectral functions at the edge of valence band along L and $\Sigma$ high symmetry points for \replaced[]{$\mathrm{Sn_{0.994}Te}$}{$\mathrm{SnTe}$} (a) without dopants; (b) with 2\% Zn; (c) with 4\% Zn; (d) with 8\% Zn. BSFs are plotted in logarithmic scale, with values higher than 300 atomic units marked as black. Panels (e-p) contain plots of BSF  near the center of the last valence band at given $\vb{k}$-points, where panels (e-h), (i-l), and (m-p) correspond to L, $\mathrm{k_s}$ and $\Sigma$ respectively. Here points represent BSF values, while solid lines are the fitted Lorentzian function. Calculated value of carrier lifetime is given at each panel.
    Representative $\rm k_s$ point was chosen in the middle between L$-\Sigma$ path, closely to the bottom of the band in that path (marked at the top row as dotted line). Each column corresponds to the same concentration, e.i., panels (e, i and m) are plotted for $\mathrm{Sn_{0.994}Te}$, (f, j and o) for $\mathrm{Sn_{0.974}Zn_{0.02}Te}$, etc. }
    \label{fig:004_bsf_ls_vb}
\end{figure*}

\subsection{Spectral functions}

Figure~\ref{fig:003_bsf_full} presents the Bloch spectral density functions along high-symmetry directions in the form of two-dimensional projections, where the color corresponds to the BSF value. 
In the case of Sn$_{0.994}$Te [Fig.~\ref{fig:003_bsf_full}(a)] one can see relatively sharp and well defined bands, where only slight smearing is visible. This smearing  is attributed to the vacancies at the Sn site. 
The introduction  of the Zn impurity into the system [Fig.~\ref{fig:003_bsf_full}(b,c)] drastically modifies the electronic bands, resulting in substantial smearing, which is further enhanced with increased doping. Moreover, in the conduction band, new features appear, which will be discussed further below. 
This is especially evident near the conduction band edge, in the vicinity of L point, at energies corresponding to the PDOS peak of Zn, i.e. at approximately 0.5 eV. 
Visible smearing, apart from the both peak in partial DOS of Zn and growth of the total DOS hump, is another piece of evidence for the resonant character of the impurity in fashion similar to that reported for other resonant impurities, such as PbTe:Tl or SnTe:In \cite{Wiendlocha2013FermiCalculations,Wiendlocha2018Thermopower,Misra2020BandLevel}. 
Concerning the character of RL, no isolated impurity band is present, as the states of the impurity and the host hybridize, creating ''clouds'' of states near the original bands. This is in complete opposite to the dispersionless spectral functions of 3d states of Zn near $-6$ eV, which are strongly localized.

Figure \ref{fig:003_bsf_full}(d) presents the unfolded band structure of the $\mathrm{Sn_{26}ZnTe_{27}}$ supercell (corresponding to SnTe with 3.7\% Zn), which can be compared to SnTe doped with 4\% Zn  [Fig.~\ref{fig:003_bsf_full}(b)]. 
In supercell calculations, Sn vacancies are absent, and thus $E_F$ is slightly shifted towards the band gap. To be consistent with BSFs, the same color map was chosen for the intensity of unfolded bands. 
The band unfolding successfully reproduced the non-trivial modifications arising from the existence of Zn impurity. This is particularly visible at the L point above the Fermi level, where bands appear to split (see below). 
Band unfolding also reproduced the flat 3d Zn bands, with their position slightly shifted towards lower energies.

Given that the last valence band has a crucial impact on the thermoelectric properties of $p$-type SnTe, the analysis of its shape and band smearing is of utmost importance.  
Figures~\ref{fig:004_bsf_ls_vb}(a) to~\ref{fig:004_bsf_ls_vb}(d)
contain magnification of the VB between the points L and $\Sigma$ for selected concentrations. Although the RL falls within the conduction band, Zn also has a strong influence on the valence band. Alloy scattering leads to the broadening of the last VB, which is especially visible for higher Zn concentrations. Additionally, no indication of RL-like distortion in VB is present in contrast to, e.g. SnTe:In or PbTe:Tl  \cite{Wiendlocha2021ResidualSemiconductors,Parashchuk2021HighSoftening}.
In Figs. \ref{fig:004_bsf_ls_vb}(e) to \ref{fig:004_bsf_ls_vb}(p), the BSFs are plotted for L, $\Sigma$ and $\mathrm{k_s}$ [located half of the distance L$-\Sigma$ (0.4375,0.4375,0.25)${2\pi}/{a}$], where the local minimum of the band is present.
Here, the shape of the BSFs highly resembles the Lorentz function as for a non-resonant case. 
The only small deviations are visible in   
Figs.~\ref{fig:004_bsf_ls_vb}(k),  \ref{fig:004_bsf_ls_vb}(l) and  \ref{fig:004_bsf_ls_vb}(p)
where Lorentzian-like BSFs are slightly asymmetrical, with ''shoulder''-like slope.
However, despite being slightly blurred, the band itself near that $\vb{k}$-point is well established considering very high values of BSF. 
As the Lorentzian shape is retained, the FWHM $\Delta$ of this function is directly related to the lifetime of the electronic state $\tau$ with the formula: $\tau=\hbar/\Delta$ \cite{Gordon1981OnAlloys}.  
For the case with 8\% Zn, with a fitted Lorentz function, we obtain: $\tau\approx3.5 \times 10^{-13}$ s, which is a noticeably high value considering the amount of doping and differs only by one order of magnitude compared to the undoped sample [Fig.~\ref{fig:004_bsf_ls_vb}(e)]. Taking into account the bottom of this part of VB, the fitting results in a much greater decrease in lifetime where the values of $\tau$ range between $\tau \approx 1.7\times 10^{-13}$ s for $\mathrm{Sn_{0.994}Te}$ and  $7.6 \times 10^{-15}$~s for $\mathrm{Sn_{0.914}Zn_{0.08}Te}$. Similar effects are observable near the $\Sigma$ $\vb{k}$-point.

\begin{figure}
    \centering
    \includegraphics[width=0.5\textwidth]{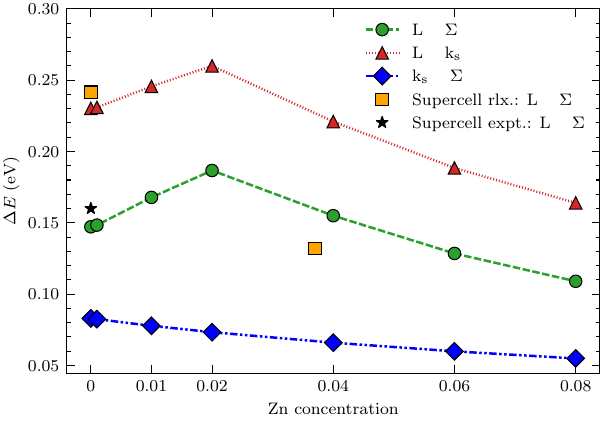}
    \caption{Energy offset $\Delta E$ between last VB maxima at L and $\Sigma$ high symmetry points and at $\mathrm{k_s}$ point as a function of Zn concentration. \added[]{For supercells without Zn, the results are given for calculations using experimental and relaxed crystal structure.}}    
    \label{fig:005_ls_band_enrgy_offset}
\end{figure}

\begin{figure*}
    \centering
    \includegraphics[width=\textwidth]{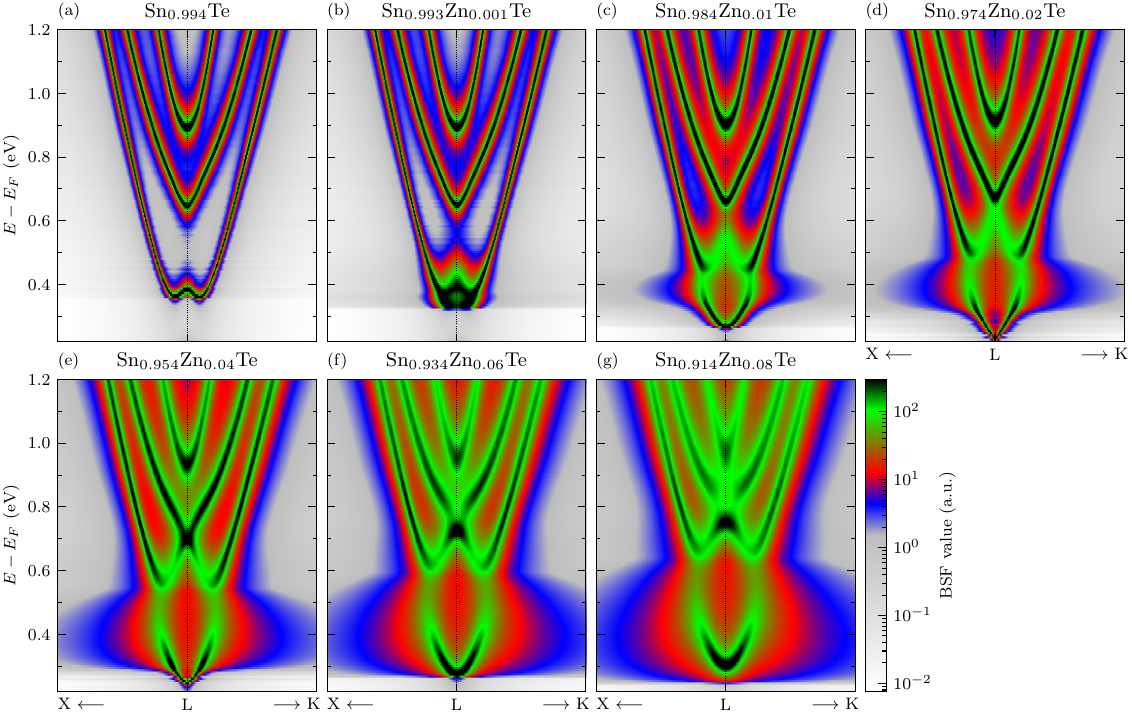}
    \caption{ (a-g) Two-dimensional Bloch spectral function near minimum of the conduction band in vicinity of the L point, alongside the X$-$L$-$K path. Colors in those panels are in logarithmic scale, with black color corresponding to values higher than 300 atomic units.   }
    \label{fig:006_XLK_VB}
\end{figure*}
Although smearing of the VB in L$-\Sigma$ might contribute to thermopower enhancement, the DOS results suggest that RL is not directly responsible for this due to its location inside CB. 
This leads to the question: what else can be the source of this improvement? The answer might be the shape of the VB, as while generally (Figure~\ref{fig:004_bsf_ls_vb}) it remains mostly unmodified, the energy offset ($\Delta E_{\text{L}-\Sigma}$) between the valence band maxima along L$-\Sigma$ direction is changing. This is shown in Fig.~\ref{fig:005_ls_band_enrgy_offset}, where the energy offset between the two band maxima and a bottom of the band at $\mathrm{k_s}$  is plotted with respect to the Zn concentration. 
In addition to $\Delta E$ obtained from the BSFs, the band offset between the last valence band maxima was also calculated from the unfolded dispersion relation for the supercell\added[]{s}. In case of a BSFs, values of the band centers at given points were estimated by the position of the peak of the fitted Lorentzian. For the unfolded bands, the position of the center of the band was chosen as the energy at the maximum intensity value. 
Here we observe that KKR-CPA calculations predict a peculiar, non-monotonic evolution, where initial energy offset $\Delta E_{\text{L}-\Sigma}$ (underestimated compared to the experiment) increases, up to 0.19 eV at 2\% Zn, and further doping leads to desired band convergence as at 8\% Zn $\Delta E_{\text{L}-\Sigma}$ is around 0.11 eV. Similar behavior can be seen for $\Delta E_{\text{L}-\mathrm{k_s}}$, while  $\Delta E_{\mathrm{k_s}-\Sigma}$ decreases in a more linear fashion. 
The result\added[]{s} from the supercell\added[]{s} confirm\replaced[]{}{s} a general trend of converging bands,
and the numerical differences between KKR-CPA and supercell\added[]{s} can be explained by the difference in methods and the single configuration studied \added[]{as well as the difference in lattice parameter in the case of relaxed variant for 0\% Zn (1.3\% larger $a$)}.
\replaced[]{If an undoped supercell with experimental lattice parameter is considered and n}{N}ear 4\% Zn\added[]{, the} results from both methods are in agreement and generally doping with Zn appears to introduce L$-\Sigma$ band convergence. The origin of this non-monotonic behavior will be discussed later.

\replaced[]{Regarding the band convergence - c}{C}omparing results for pristine SnTe found in the literature, our KKR-CPA + LDA calculations underestimate the band offset (experimental value of around 0.3 eV  \cite{Rogers1968ValenceSnTe} and theoretical in the range of $0.25-0.36$ eV \cite{Dong2016First-principlesDoped-SnTe,Bhat2019Zn:Thermoelectrics,Wang2024,Zhang2021EnhancedDefects}), while the pseudopotential  approach  \added[]{for the relaxed structure} yields a closer value ($0.24$ eV). 
For Zn-doped variants, the values of $\Delta E_{\text{L}-\Sigma}$ from the literature can be found only for SnTe with $\approx4$\% of Zn. In such case, calculations yield $\Delta E_{\text{L}-\Sigma}=0.12-0.15$~eV which slightly overstimates the band offset with respect to the reported values ($0.04-0.07$ eV) \cite{Dong2016First-principlesDoped-SnTe,Wang2024}.
In addition to the band convergence, both the decrease in $\Delta E_{\text{L}-\mathrm{k_s}}$ and  $\Delta E_{\mathrm{k_s}-\Sigma}$ means that the curvature of this band decreases, resulting in an increase in the effective mass of carriers. Those three effects, i.e., band convergence, flattening, and smearing of the VB, are responsible for the enhancement of $S$ in SnTe:Zn, which we further demonstrate using the transport calculations below.


Earlier we mentioned that the addition of Zn affects the CB in a non-trivial fashion. This effect is shown in Fig.~\ref{fig:006_XLK_VB}, where BSFs near the L point (in the direction X$-$L$-$K) are presented for each concentration studied. In panel Fig.~\ref{fig:006_XLK_VB}(a) one can see that LDA predicts the CB minimum to be slightly shifted from the L point itself. 
This feature is in agreement with previous reports \cite{Misra2020BandLevel,Misra2022Influence/math,Wiendlocha2021ResidualSemiconductors,Littlewood2010}.  
The addition of a small amount (0.1\%) of Zn [Fig.~\ref{fig:006_XLK_VB}(b)] suppresses this trait, since due to the presence of Zn impurity states, significant smearing and band reconstruction begin around the L point. 
For 1\% Zn one can clearly observe that the original CB of $\mathrm{Sn_{0.994}Te}$ splits into two parts, one of which becomes a new CB with a minimum directly at point L, and the other moves towards the second, higher conduction band. 
With sufficiently high doping ($\geq 2\%$) these bands join at the L point, and for even higher doping with the third conduction band [Fig.~\ref{fig:006_XLK_VB}(g)].
The bottom of the CB remains at the L point, connected with the higher conduction bands through a strongly smeared bridges.

\begin{figure*}
    \centering
    \includegraphics[width=\textwidth]{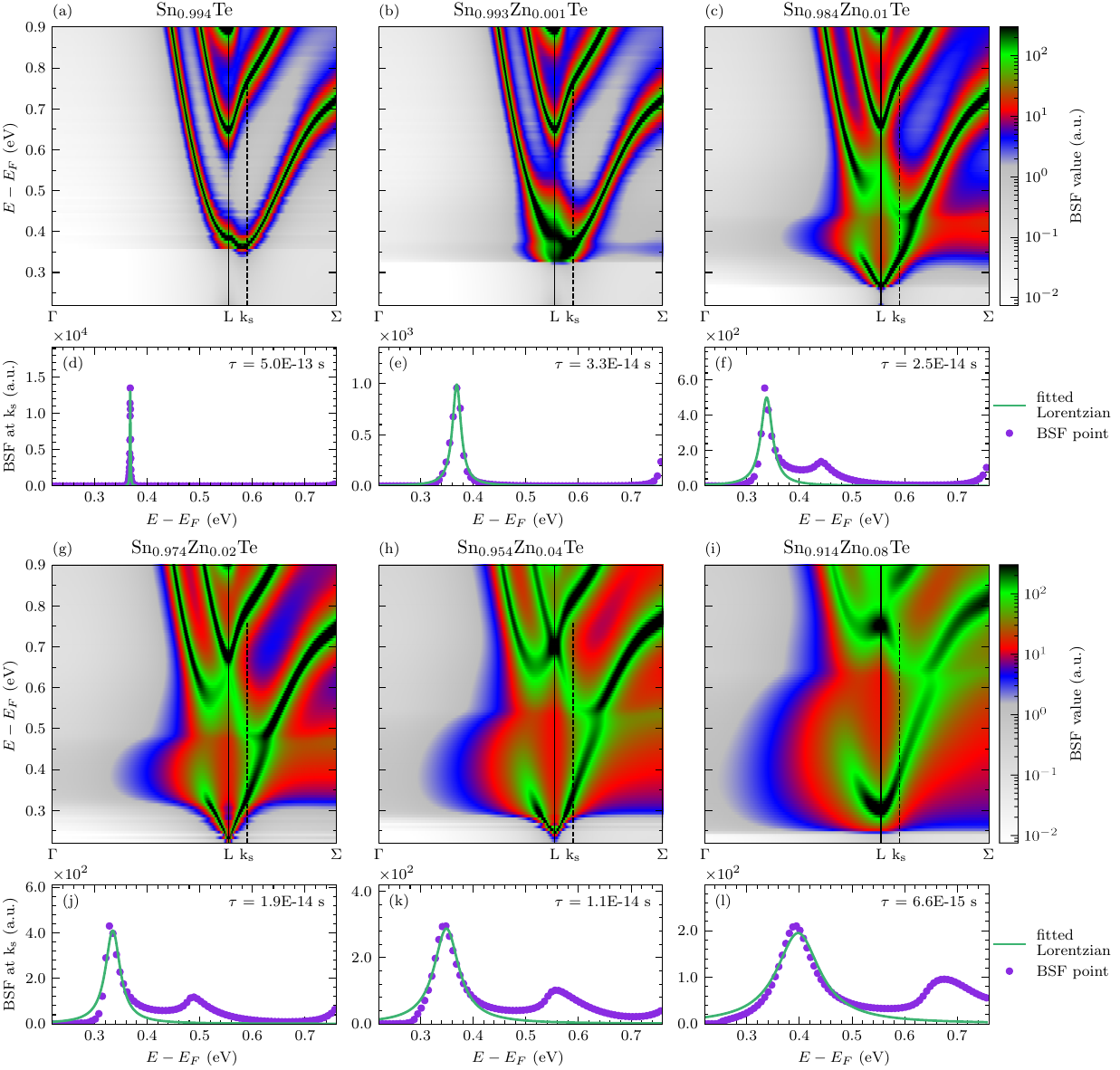}
    \caption{(a-c) and (g-i) Two-dimensional Bloch spectral function near minimum the conduction band along the $\Gamma$$-$L$-$$\Sigma$ path. Colors in those panels are in logarithmic scale, with black color corresponding to values higher than 300 atomic units. Below each of these panels (d-f and j-l), BSFs are plotted for $\mathrm{k_s}\approx( 0.479,0.479,0.414) {2\pi}/{a}$ that is about $1/6$ of the L$-$$\Sigma$ distance. Here,  points represent calculated BSFs, and solid lines are a fit to the Lorentzian function. Estimate of the carrier lifetime based on the fitted peak FWHM is shown in each figure. While for low concentrations (a-c)  the Lorentzian function describes well shape of BSFs, above 2\% of Zn description of the BSF using a single Lorentz function is no longer possible.   }
    \label{fig:007_GLS_VB}
\end{figure*}

This ''disconnecting'' tendency is weaker in the lowest CB part in the L$-\Sigma$ direction --
in Fig.~\ref{fig:007_GLS_VB} two-dimensional projections of BSFs in the $\Gamma$$-$L$-$$\Sigma$ path are shown and band reconstruction takes place, new band connections are formed, but keeping the high intensity connection from the CB bottom at L to $\Sigma$, leaving the robust  L$-\Sigma$ band.
In addition, in the lower panels the calculated energy-dependent BSFs at a selected
$\mathrm{k_s}$ point
are shown to analyze their shape
[Figs.~\ref{fig:007_GLS_VB}(d) to \ref{fig:007_GLS_VB}(f) and Figs.~\ref{fig:007_GLS_VB}(j) to \ref{fig:007_GLS_VB}(l)].
The coordinates of $\mathrm{k_s}$ are (0.479,0.479,0.414)${2\pi}/{a}$ and this point 
was chosen to illustrate the 
shape of BSF. 
Both small $\tau$ values as well as non-Lorentzian shape is a typical behavior for resonant impurities \cite{Wiendlocha2021ResidualSemiconductors,Wiendlocha2013FermiCalculations,Gyorffy1979FirstAlloys}.
The main cause for observed band splitting in CB upon doping is described in the following section.

\added[comment=R1Q1 R1Q3]{The presented band structure of Zn-doped SnTe (both obtained via KKR-CPA and supercell approach) contrasts with the results available in the literature \cite{Zhang2021EnhancedDefects,Bhat2019Zn:Thermoelectrics,Shenoy2020BiiZT/i,Dong2016First-principlesDoped-SnTe,Wang2024} in terms of effects of Zn addition on the general shape of the dispersion relation and the matter of the band gap. It is important to note though, that so far all the existing works devoted to SnTe:Zn where calculations were present showed a regular dispersion relation for supercells, thus the Zn-induced band would naturally appear alike to a localized impurity state. In works by Bhat and Shenoy \cite{Bhat2019Zn:Thermoelectrics} as well as Zhang et al. \cite{Zhang2021EnhancedDefects} after substituting one Sn atom with Zn in the supercell (Sn$_{31}$ZnTe$_{32}$ - effective concentration of ~3.1\%, comparable with Sn$_{26}$ZnTe$_{27}$ shown in this work), a new band is formed, spanning from the bottom of the conduction band to the VB and overlapping with it by over 0.2 eV, closing the band gap in the process. This is not the case here - see Fig.\ref{fig:003_bsf_full}, where a clear band gap is present. Similar behavior with floating band that closes the band gap is present in smaller supercells: 
2$\times$2$\times$1 \cite{Bhat2019Zn:Thermoelectrics} and  $\sqrt{2}$$\times$$\sqrt{2}$$\times$2 \cite{Shenoy2020BiiZT/i}, with a much larger Zn concentration of ~6.3\%. In all of the above-mentioned publications full structural relaxation were performed, however, to the best of our knowledge  only in  work \cite{Zhang2021EnhancedDefects} a force convergence threshold was reported of a  20 meV/\AA\, which is much larger compared to 1 meV/\AA\ utilized here indicating that a reason for this discrepancy might lie in not sufficient relaxation of the unit cell. In Fig. S2(a,b)~\cite{suppl} an unfolded band structure of the Sn$_{26}$ZnTe$_{27}$ supercell is shown before and after structural relaxation. In the unrelaxed structure, the band forming bottom of the CB spans through the band gap and slightly overlaps with the VB making this system semimetallic and forming a feature that is similar as in the above-mentioned works. After a strict structural relaxation [Fig. S2(b)~\cite{suppl}] that band no longer overlaps with VB, and the band structure is more similar to that obtained via KKR-CPA method indicating the large role of structural optimization in case of the supercell calculations. Results obtained in \cite{Dong2016First-principlesDoped-SnTe} appear in better agreement in terms of the DOS as discussed previously, however from the perspective of band structure results reported in \cite{Dong2016First-principlesDoped-SnTe} are more in favor of Zn forming an isolated and localized impurity band, which has been disproved in our work by analysis of the BSFs. Finally, in \cite{Wang2024} addition of Zn in SnTe appears as an emergence of a band at the bottom of the conduction band that overlaps with the existing bands which is in good agreement with the results presented in current work. Regarding the sources of discrepancies with respect to the problem of the gap closed by Zn doping, it is worth noting that the symmetry of the supercell used in calculations is correlated with the result. In Refs. \cite{Zhang2021EnhancedDefects,Bhat2019Zn:Thermoelectrics,Shenoy2020BiiZT/i} the supercells were constructed based on the conventional unit cell, resulting in a simple cubic supercell structure. The band gap was closed in those cases for the structures containing Zn. This is not the case both in our case and in Refs. \cite{Dong2016First-principlesDoped-SnTe,Wang2024} with 3×3×3 supercells based on the primitive face-centered cell, where no gap closing was observed.}

\subsection{Band inversion}

    \begin{figure*}
        \centering
        \includegraphics[width=\textwidth]{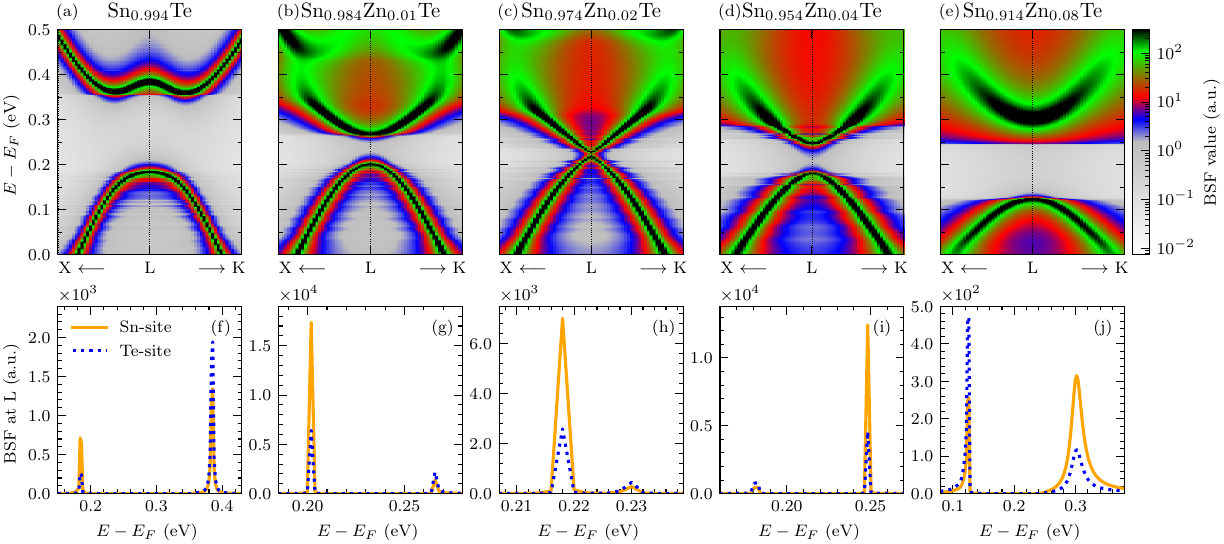}
        \caption{ (a-e) Two-dimensional Bloch spectral function near the band gap in vicinity of the L point, alongside the X$-$L$-$K path. Colors in those panels are in logarithmic scale, with black color corresponding to values higher than 300 atomic units. Panels (f-j) show BSFs at L high symmetry point with their site character projected. Each column corresponds to the same concentration, i.e. panel (f) is plotted for $\mathrm{Sn_{0.994}Te}$, (g) for $\mathrm{Sn_{0.984}Zn_{0.01}Te}$, etc.}
        \label{fig:208_bsf_gap_inv}
    \end{figure*}

As discussed earlier, we observed a non-monotonic behavior in the band convergence with Zn content (Fig.~\ref{fig:005_ls_band_enrgy_offset}), where the energy offset initially increases up to 2\% and then starts decreasing. 
Similarly, the unique band splitting at the bottom of CB (Figs.~\ref{fig:006_XLK_VB} and~\ref{fig:007_GLS_VB}) also raises questions.
To reveal the origin of this peculiar evolution, we need to take a closer look at the development of BSFs in the vicinity of the band gap with respect to the Zn content (Fig.~\ref{fig:208_bsf_gap_inv}). 

From SnTe with vacancies, the indirect band gap of $E_g\sim0.18$ eV is present between the valence band maximum (VBM) at the L point and the conduction band minimum (CBM) which has a small offset from the high symmetry point. The gap value is in very good agreement with the experiment \cite{PhysRevLett.16.1193}. As the concentration of dopant increases, the CBM formed from a splitted band is  now located at the L point and the band itself is shifted towards lower energies, decreasing the band gap. At the same time, VBM is slightly pushed to higher energies causing the non-monotonic band convergence as a side effect. 
This behavior continues until 2\% where the band gap nearly vanishes, and the bands at both edges are nearly linear, forming a Dirac cone. 
As the Zn content increases further, the band gap starts to reopen again and returns to almost the original width of $E_g\sim0.18$~eV at 8\%. This is shown quantitatively in Fig \replaced[]{S3}{S2}~\cite{suppl}, where the estimation of band gap width $E_g$ from the fitted Lorentzian peak in the vicinity of the L-point is shown. \replaced[comment=R3Q1]{Considering the inaccuracy of predicting the band gap width by LDA, we estimate that the computationally predicted critical concentration may range depending on the choice of exchange-correlation potential. KKR-CPA with LDA of Vosko et al. \cite{Vosko1980AccurateAnalysis} correctly reproduced the band gap of SnTe 0.18 eV and predicted the critical concentration for gap closing of 2\% of Zn. For GGA-PBE where the gap is underestimated in the relaxed supercell ($E_g = 0.11$ eV) a smaller critical concentration can be expected, of the order of 0.5\% - 1\%. Consequently, exchange-correlation functional which overestimates the gap, like CA LDA in Ref.\cite{Drppel2014} ($E_g = 0.3$ eV), should predict a larger critical concentration.}{}

Observed closing and reopening of the band gap indicates that due to Zn doping the system undergoes  a topological phase transition from the inverted band state of the topological insulator (TI) to a trivial state of the insulator \cite{Hsieh2012,Drppel2014}. 
SnTe has an inverted band structure: the cation (Sn) states dominate at the VBM, and the anion (Te) states dominate at the CBM, in contrast to the typical trivial band insulator, which is, e.g. PbTe.
SnTe was the first example of the topological crystalline insulator (TCI) \cite{Hsieh2012,Tanaka2012} where instead of the time-reversal symmetry as in regular TI \cite{Fu2011}, the mirror structural symmetry ensured topological protection. 
The emergence of band inversion itself can also be controlled. Increasing the lattice parameter can allow tuning the gap in SnTe and eventually turning it into a conventional insulator with a sufficiently high lattice parameter, as predicted by DFT-based studies \cite{Hsieh2012,Drppel2014}. At the same time, while other members of the IV-VI semiconductors are normal insulators, it is possible to induce a topological phase transition in most of them by applying pressure \cite{Barone2013,Liang2017,Pal2020}, temperature \cite{Dziawa2012,Wojek2014,Krizman2018} or by varying composition of SnTe-based alloys \cite{Calawa1969,Strauss1967,Gao2008,PhysRevLett.16.1193,Xu2012,Krizman2018,Tanaka2013,Lusakowski2018}. 
In addition to alloying, defect engineering, such as Cr doping, has been utilized to turn SnTe into a Weyl semimietal with the transition controlled by varying the concentration \cite{Pham2019}. Similarly, sufficiently high In doping is capable of turning SnTe into a Dirac semimetal phase \cite{Schmidt2020}. However, to the best of our knowledge, no report on Zn-induced topological phase transition has been reported so far. 
One work reported using band inversion to improve the thermoelectric capabilities of SnTe \cite{Xie2020}, opening a new path for further improvement.

Moving back to the results, as mentioned, in contrast to other members of the IV-VI semiconductor family, SnTe has an intrinsically inverted band structure. In PbTe, the band at the CBM has L$_6^-$ symmetry and domination of a cation (Pb) p-states, while at the VBM it's L$_6^+$ with Te-p-states domination; for SnTe the order is reversed \cite{Gao2008,PhysRevLett.16.1193,Koumoulis2015}. 
The origin of this difference lies in a stronger Sn 5s-Te 5p hybridization in SnTe due to the higher energy of the 5s orbital. In the case of PbTe due to stronger relativistic effects, the orbital Pb-6s has a lower energy, and hybridization with Te 5p is weaker and inversion is absent \cite{Ye2015}. 

Figure ~\ref{fig:208_bsf_gap_inv}(f), where the BSFs are plotted at the L point with their site character projected: Sn-site (4a) and Te-site (4b), confirms this inversion. We observe that VBM has the dominant contribution from the Sn-site, while in CBM, the contribution from the Te site is higher, in agreement with previous reports \cite{Hsieh2012,Gao2008,Drppel2014,Dziawa2012,QueralesFlores2020}. 
Now when Zn is added to the system, the inversion along with the band gap gradually vanishes, for 4\% and 8\% of Zn [Fig.~\ref{fig:208_bsf_gap_inv}(i,j)] when the gap is reopened, the trivial band ordering is restored.

\begin{figure}
    \centering
    \includegraphics[width=0.5\textwidth]{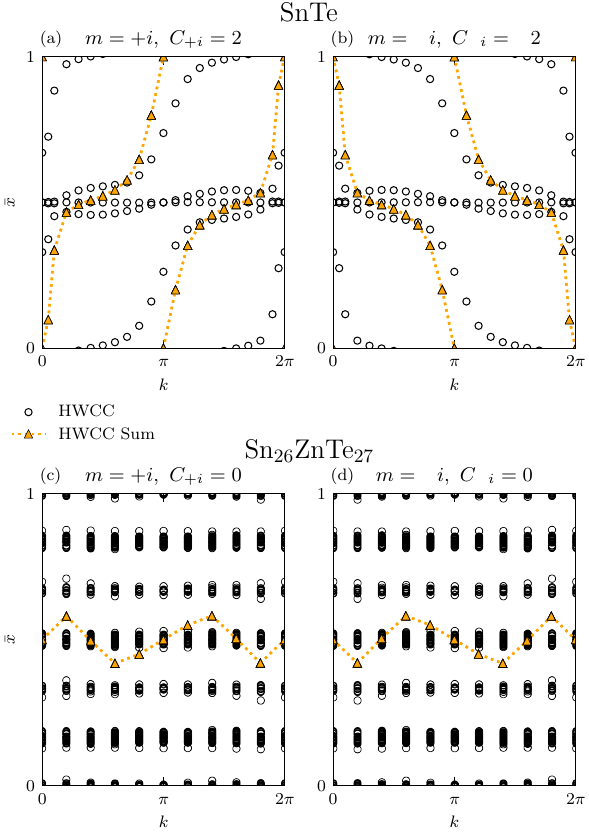}
    \caption{\added{The calculated hybrid Wannier charge centers (HWCCs) marked with circles and their sum (triangles) for the $m\pm i$ eigenstates in the mirror plane for (a,b) SnTe, (c,d) Sn$_{26}$ZnTe$_{27}$. Calculated mirror Chern number for the SnTe is $n_M=2$, while in the case of Sn$_{26}$ZnTe$_{27}$, the value is $n_M=0$, indicating topologically trivial phase. }}
    \label{fig:hwcc}    
\end{figure}

\replaced[comment=R3Q4]{To further confirm the emergence of topological phase transition with doping, we also demonstrate the inversion of band parities at L point in the calculated supercells for the doped and pristine system (See Fig. S4 in the Supplemental Material~\cite{suppl}). Moreover, in the case of both, SnTe and Sn$_{26}$ZnTe$_{27}$ we have calculated the hybrid Wannier charge centers (HWCCs) in the [110] mirror plane for the $m\pm i$ eigenstates as well as the mirror Chern number $n_M=(C_{+i}-C_{-i})/2$ \cite{Teo2008}, where $C_{\pm i}$ are individual Chern numbers for each of the two subspaces separated according to their mirror eigenvalues $m=\pm i$, respectively. The results are shown in Fig.~\ref{fig:hwcc}. We start with the case of pristine SnTe, which serves as a reference point. As implied by the double winding of the HWCCs' sum in Fig.~\ref{fig:hwcc}(a,b) we obtain $C_{\pm i}=\pm2$ for each of subspaces, thus the resulting mirror Chern number is $n_M=2$ indicating that SnTe is topologically non-trivial - a topological crystalline insulator - in agreement with previous calculation \cite{Hsieh2012}. In contrast, for the Sn$_{26}$ZnTe$_{27}$ we observe a different behavior, as both  HWCCs spectrums [Fig.~\ref{fig:hwcc}(c,d)] possess multiple gaps and in addition, no winding is present in the sum of HWCCs. The value of each $C_{\pm i}$ and $n_M$ is thus 0 further confirming the presence of a topologically trivial phase in this system.}{}

It is worth noting that the influence of Zn on the band inversion is stronger than that of the lattice parameter $a$. In case of SnTe, the transition to trivial insulator can be caused by increasing the lattice parameter sufficiently, with the transition predicted to appear between 1.7\% to 4.1\% larger $a$ over the optimal lattice constant \cite{Hsieh2012,Drppel2014}. With Zn doping, $a$ \textit{decreases} by 0.5\textperthousand(1\textperthousand) for $\mathrm{Sn_{0.974}Zn_{0.02}Te}$ ($\mathrm{Sn_{0.954}Zn_{0.04}Te}$), which should widen the inverted band gap assuming that the doping would only affect the lattice parameter \cite{Hsieh2012,Drppel2014}.

\subsection{Transport properties}

 \begin{figure}
    \centering
    \includegraphics[width=0.5\textwidth]{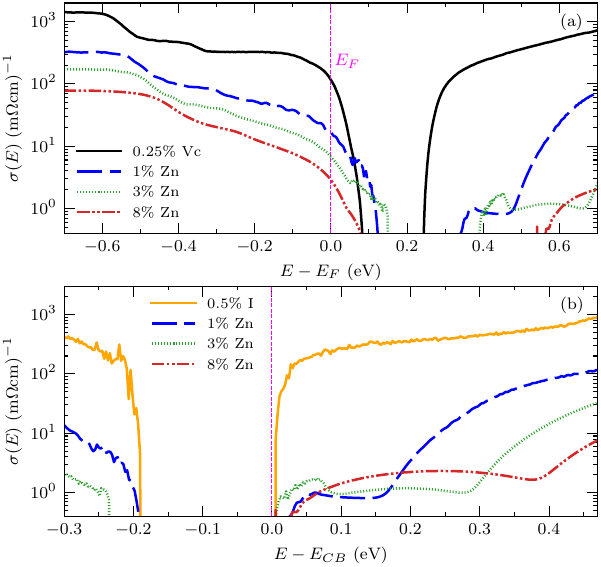}
    \caption{Energy-dependent transport function $\sigma(E)$ of Sn$_{0.994-x}$Zn$_x$Te for $x=0.01,0.03, \text{ and } 0.08$. Panel (a) focuses on the valence band with added $\sigma(E)$ for $\mathrm{Sn_{0.9975}Te}$ as a $p$-type reference system. In panel (b) zoom on the conduction band is presented, with the transport function for $\mathrm{SnTe_{0.995}I_{0.005}}$, serving as a reference $n$-type case. Here, the energy scale is shown with the reference to the edge of conduction band.}
    \label{fig:transport_sigma}    
\end{figure}

\begin{figure*}
    \centering
    \includegraphics[width=\textwidth]{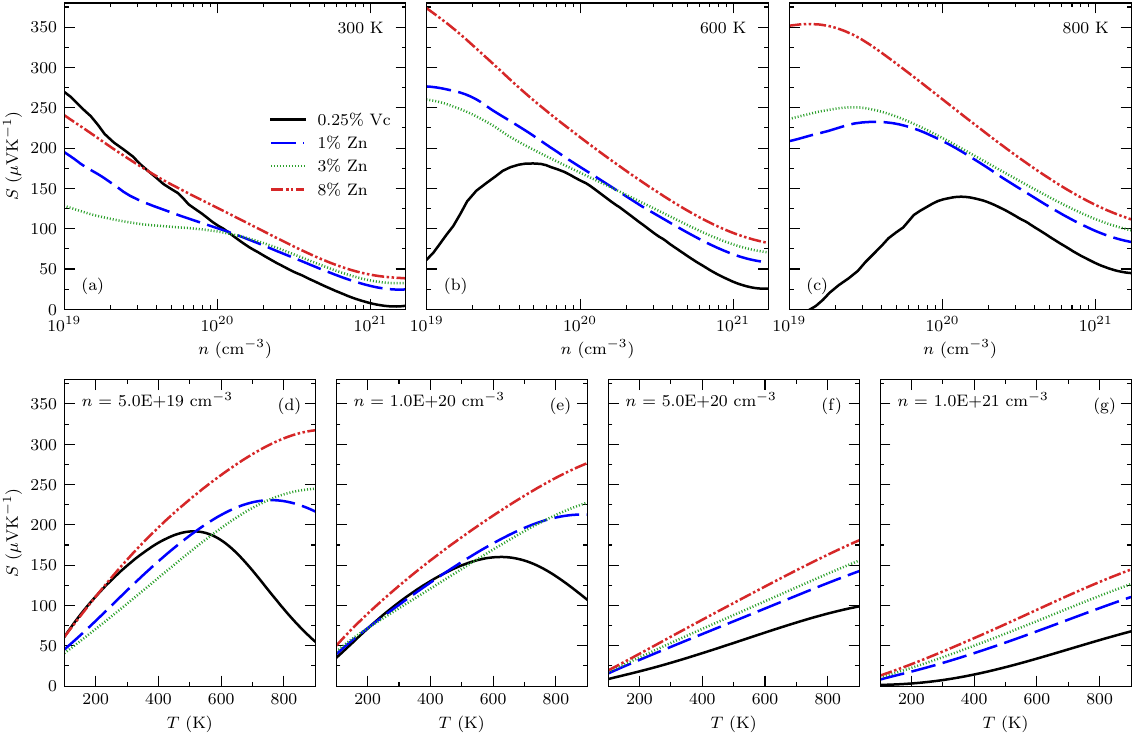}
    \caption{Transport properties of $p$-type SnTe doped with 1, 3 and 8\% of Zn as well as $\mathrm{Sn_{0.9975}Te}$. Panels (a-c) show calculated Seebeck coefficient as a function of hole concentration for 300, 600 and 800 K. In panels (d-g) temperature-dependent $S$ is presented for selected carrier concentrations.}
    \label{fig:transport_summary_ptype}
\end{figure*}
\begin{figure}
    \centering
    \includegraphics[width=0.5\textwidth]{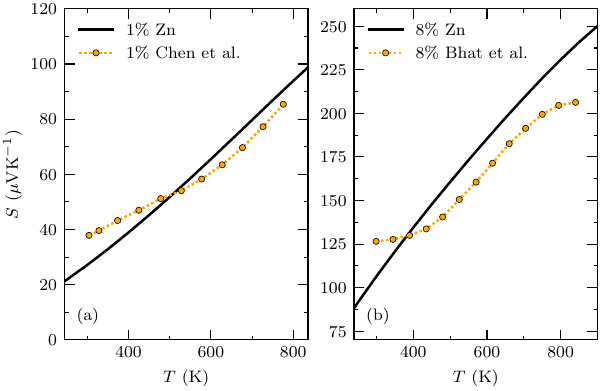}
    \caption{Calculated temperature-dependent Seebeck coefficient with experimental data from literature shown as a reference for (a) 1\%  \cite{Chen2018BandZn-doping} and  (b) 8\% Zn-doped SnTe \cite{Bhat2019Zn:Thermoelectrics}. Theoretical results were obtained for concentrations corresponding to the experimental samples ($1.1\times10^{21}$ cm$^{-3}$ for $\mathrm{Sn_{0.99}Zn_{0.01}Te}$~\cite{Chen2018BandZn-doping} and $1.6\times10^{20}$ cm$^{-3}$ for $\mathrm{Sn_{0.96}Zn_{0.08}Te}$ ~\cite{Bhat2019Zn:Thermoelectrics})}
    \label{fig:seebeck_vs_expt}
\end{figure}
To study the influence of Zn doping on thermoelectric performance, calculations of transport properties were performed employing the combination of the KKR-CPA method and the Kubo-Greenwood formalism. This approach has been successfully applied for studies of transport properties in various thermoelectric materials
\cite{Wiendlocha2018Thermopower,Wojciechowski2020,Banhart1995,PhysRevB.29.4217,Banhart1995,Wiendlocha2018Thermopower,Vernes2003,Li2025,Pryga2025}. Within this method, the energy-dependent electrical conductivity function of the ground state ($T=0$ K) $\sigma(E)$ (called further the transport function) can be obtained in a parameter-free way. 
In a case where impurity-induced electron scattering is a dominant mechanism, based on $\sigma(E)$ the Seebeck coefficient may be calculated using full Fermi integrals:
\begin{equation}
    S=-\dfrac{1}{eT} \dfrac{L^{(1)}}{L^{(0)}},\label{eq:seebeck}
\end{equation}
where:
\begin{equation}
    L^{(n)}=\int dE \qty(-\pdv{f}{E})(E-\mu)^n\sigma(E).\label{eq:ln}
\end{equation}
The $e$, $f$, and $\mu$ stand for electronic charge, Fermi-Dirac distribution function, and chemical potential, with the latter calculated at each given position of $E_F$ and temperature.
This calculation goes beyond the usual constant-relaxation-time approximation and allows us to observe the resonant enhancement of thermopower~\cite{Wiendlocha2018Thermopower}. Electron-phonon scattering is, however, not taken into account and the temperature effects are related to the smearing of the Fermi-Dirac distribution.

\begin{figure*}
    \centering
    \includegraphics[width=\textwidth]{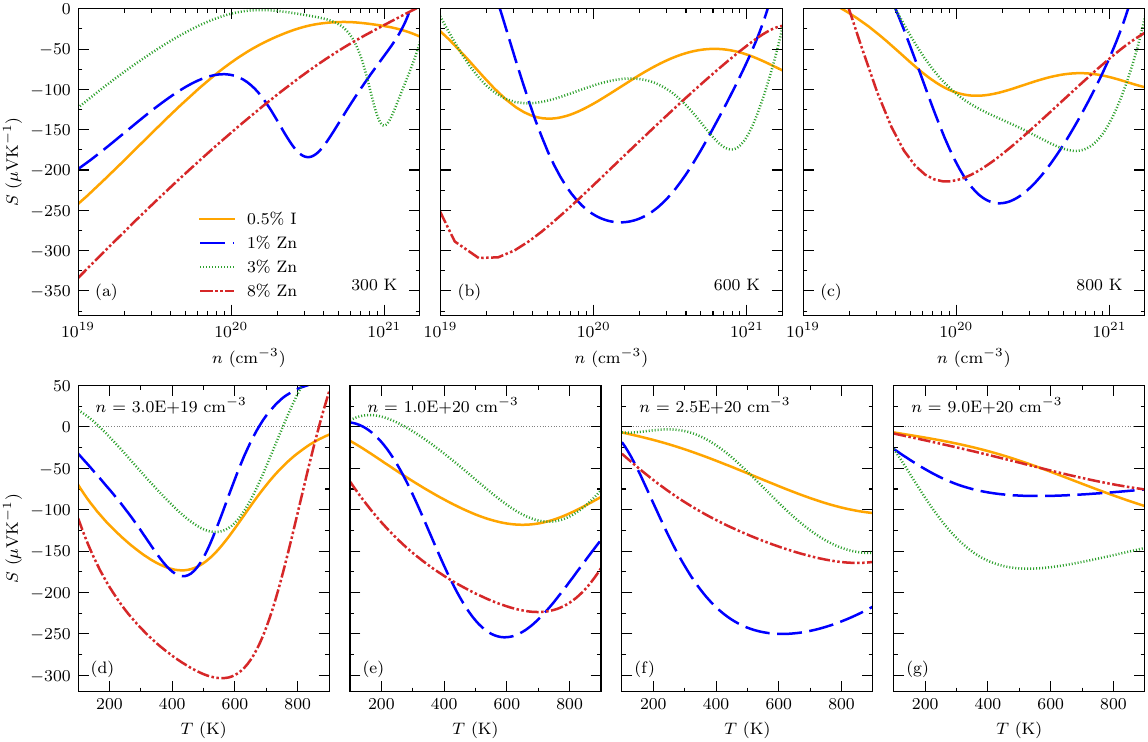}
    \caption{Transport properties of $n$-type SnTe doped with 1, 3 and 8\% of Zn as well as  $\mathrm{SnTe_{0.995}I_{0.005}}$. Panels (a-c) show calculated Seebeck coefficient as a function of electron concentration for 300, 600 and 800 K. In panels (d-g) temperature-dependent $S$ is presented for selected carrier concentrations.}
    \label{fig:transport_summary_ntpe}
\end{figure*}

In Figs.~\ref{fig:transport_sigma},~\ref{fig:transport_summary_ptype} and~\ref{fig:transport_summary_ntpe} the summary of transport properties calculation for SnTe doped with 1, 3 and 8\% of Zn is shown. 
In addition, two variants: $\mathrm{SnTe_{0.995}I_{0.005}}$ and SnTe with 0.25\% of Sn vacancies are also presented as reference results for the $n$-type and $p$-type materials, respectively. 
In both cases, the lattice parameters are kept the same as in $\mathrm{Sn_{0.994}Te}$. Experimental band gaps were used for the transport calculation, i.e. 0.43 eV for SnTe with 8\% Zn and 0.23 eV for 3\% Zn \cite{Bhat2019Zn:Thermoelectrics}. For the remaining cases with lower impurity concentrations (1\% Zn, 0.5\% I and 0.25\% Vc) we used the value of $E_g = 0.18$~eV, which corresponds to the experimental band gap for a pristine SnTe \cite{PhysRevLett.16.1193}.

Figure~\ref{fig:transport_sigma} presents the energy-dependent transport function $\sigma(E)$ for all studied samples. The overall shape of the transport function is similar to the density of states (Fig.~\ref{fig:002_dos_compilation}), albeit with some minor differences.

Starting with the transport function for $p$-type materials [Fig.~\ref{fig:transport_sigma}(a)], in all cases the $\sigma(E)$ function gradually increases for the energy states located deeper within the valence band, with three small plateaus visible in the reference system (0.25\% Vc) at $-0.1$, $-0.4$ and $-0.6$ eV. 
Considering the Zn-doped SnTe, significant decrease of the $\sigma(E)$ with increasing Zn concentration is observed in the whole energy range at both valence and conduction band. This decrease in conductivity with Zn content was also observed in some of the experimental works \cite{Bhat2019Zn:Thermoelectrics,Wang2024}. The cause of this behavior becomes clear if we consider the Drude model: $\sigma=\frac{n e^{2} \tau}{m^{*}}$, where $n$ and $m^{*}$ correspond to the carrier concentration and effective mass, respectively. The decrease in electronic lifetime with doping (see Fig.~\ref{fig:004_bsf_ls_vb}) directly decreases the transport function. Similarly, flattening of the valence band contributes to the reduction of $\sigma$ due to the increase $m^{*}$ with the Zn content. 

Thermopower, calculated using Eq.~\eqref{eq:seebeck}, is plotted as a function of carrier concentration in Figures~\ref{fig:transport_summary_ptype}(a-c) for 300, 600 and 800 K. In addition, Figs.~\ref{fig:transport_summary_ptype}(d-g) present the temperature-dependent Seebeck coefficient at selected carrier concentrations. 
Starting with the reference system (variant with 0.25\% Vc only), we can see that at room temperature, the thermopower drops very quickly with increasing carrier concentration. 
Generally, in all temperatures and carrier concentrations, Zn doping increases the thermopower. Only at 300 K and around $10^{19}$~cm$^{-3}$ it is predicted that the pristine SnTe has a higher $S$.
Considering higher temperature range (600, 800 K), among all other $p$-type samples, this one is affected by the bipolar effect in the strongest manner [as indicated by Fig.~\ref{fig:transport_summary_ptype}(d, e)], even achieving a negative Seebeck coefficient near $10^{19}$ cm$^{-3}$ at 800 K.

Moving on to the transport properties of SnTe:Zn, for both samples with 1\% and 3\% of Zn, the thermopower remains at a similar level, with visible differences observed only in the lower carrier concentration range [Fig.~\ref{fig:transport_summary_ptype}(a-d)]. 
Above $10^{20}$ cm$^{-3}$ and in the temperature range considered [Fig.~\ref{fig:transport_summary_ptype}(e-g)] the differences in $S$ are not greater than $15$ $\mu$V/K with the variant of 3\% Zn performing slightly better. 
Finally, \replaced[comment=R2Q2]{$\mathrm{Sn_{0.914}Zn_{0.08}Te}$}{$\mathrm{Sn_{0.914}Sn_{0.08}Te}$} surpasses in terms of Seebeck coefficient all remaining variants in the majority of the carrier concentration range [Fig.~\ref{fig:transport_summary_ptype}(d-g)], achieving over 250 $\mu$V/K for $10^{20}$ cm$^{-3}$ at 800 K.
A small decrease in $S$ due to the bipolar effect can be observed for Zn-doped variants up to $n<5\times 10^{19}$ cm$^{-3}$. 
However, due to the high carrier concentration of the SnTe samples in general (range of $10^{20}$ to $10^{21}$ cm$^{-3}$ \cite{Moshwan2017EcoFriendlyChallenges,Chen2018BandZn-doping,Bhat2019Zn:Thermoelectrics}), this effect is not a serious limitation. 
Our calculations show a general improvement of Seebeck coefficient in \textit{p}-type SnTe through zinc doping for high concentrations of Zn.

Comparison of the calculated and experimental thermopower as a function of temperature is given in Fig.~\ref{fig:seebeck_vs_expt}. 
Direct comparison was only possible for SnTe with 1\% and 8\% of Zn, where the carrier concentration in samples was reported: $1.1\times10^{21}$ cm$^{-3}$ for $\mathrm{Sn_{0.99}Zn_{0.01}Te}$~\cite{Chen2018BandZn-doping} and $1.6\times10^{20}$ cm$^{-3}$ for $\mathrm{Sn_{0.96}Zn_{0.08}Te}$ ~\cite{Bhat2019Zn:Thermoelectrics}. 
As one can see, agreement between our calculations and the experiment is rather good in both cases. 
In the first case [Fig.~\ref{fig:seebeck_vs_expt}(a)], an underestimate of the theoretical $S$ is observed only at temperatures below $400$~K. Taking into account the variant with 8\% Zn, our calculations overestimate the thermopower by about $15$ $\mu$V/K in the medium to high-temperature range (difference up to 15\%). 
The primary reason for the observed overestimation is probably the neglect of electron-phonon scattering in our computations; this effect will reduce conductivity, flatten the $\sigma(E)$ curve and lower the thermopower.

We should mention that our calculations predict an overestimation of the thermopower at low to medium temperatures in pristine SnTe with respect to the experiment. The predicted room-temperature $S$ is close to 65~$\mu$V/K, while experimental works report thermopower in the range of 10-25 $\mu$V/K \cite{Bhat2019Zn:Thermoelectrics,Zhang2013HighSnTe,Zhou2014OptimizationBand}. Similarly, at 600~K, the experimentally measured values are overestimated by about 55~$\mu$V/K. This discrepancy is caused by insufficient energy separation between the last two VB maxima in the L$-\Sigma$ direction, as discussed in the previous section. Similar thermopower was reported in other theoretical works for SnTe \cite{SINGH2010}, and the underestimated L$-\Sigma$ distance also affected the calculated thermopower in PbTe~\cite{Wiendlocha2018Thermopower}.
However, the predicted enhancement in Zn-doped SnTe can be treated as valid considering that all of the calculations were performed in the same way, possibly leading to a comparable and systematic offset among all variants.


To discuss the effect of Zn doping on $n$-type transport properties, SnTe doped with 0.5\% I was chosen as the reference material. Introducing iodine to SnTe causes almost solely a rigid shift of the Fermi level which is pushed into the CB (See Fig. \replaced[]{S5}{S3}~\cite{suppl}) as I substituting Te acts as an electron donor. Apart from that, no substantial modifications to both DOS [Fig. \replaced[]{S5}{S3}(a)~\cite{suppl}] or BSF [Fig. \replaced[]{S5}{S3}(b)~\cite{suppl}] are present when compared to pristine SnTe with vacancies [Figs.~\ref{fig:001a_dos_comp}(a) and \ref{fig:003_bsf_full}(a)], making it an ideal reference point for discussing effects of Zn in $n$-type material.

Unfortunately, due to the intrinsically large amount of Sn-vacancies synthesizing $n$-type SnTe remains an elusive task, and so far no successful attempts have been reported. However, it is possible to achieve this system through heavy alloying (40\%) with Pb followed by I or Br doping \cite{Pang2021,Guo2022,Wang2025ntype}, emphasizing the need for further effort in this direction.

The transport function of SnTe$_{0.995}$I$_{0.005}$ appears to be similar to that of $\mathrm{Sn_{0.9975}Te}$ for the $p$-type, as shown in Fig.~\ref{fig:transport_sigma}. The calculated thermopower (Fig. \ref{fig:transport_summary_ntpe}) is also similar in terms of the absolute value $S$ and its temperature dependence.

Moving to the effect of Zn doping on $n$-type thermoelectric properties, the influence of Zn on the transport function in the conduction band is much more prominent, as seen in Fig.~\ref{fig:transport_sigma}(b). 
For clarity, the energy scale on that plot is presented with respect to the CB edge. The general decrease of $\sigma(E)$ is stronger compared to the case of VB for $p$-type transport. In addition, small peaks appear in the $\sigma(E)$ at the edge of the conduction band, which correspond to the resonant enhancement of DOS (Fig.~\ref{fig:002_dos_compilation}).

Fig.~\ref{fig:transport_summary_ntpe}, shows the calculated thermopower as a function of the electron concentration [Fig.~\ref{fig:transport_summary_ntpe}(a-c)], as well as temperature-dependent $S$ for selected carrier concentrations in SnTe:Zn [Fig.~\ref{fig:transport_summary_ntpe}(d-g)]. Within the electron-doped material, modifications caused by the potential resonant level in the conduction band strongly affect the Seebeck coefficient. Improvement in thermopower is more selective in contrast to $p$-type transport, as Zn-doped systems appear better than the reference system only in narrower concentration ranges, which are sensitive to temperature and are different for each of the selected Zn concentration variants.

At room temperature, for both SnTe with 1\% and 3\% of Zn, large peaks in $S$ are present near $3\times 10^{20}$ cm$^{-3}$ and $10^{21}$ cm$^{-3}$, respectively, with regard to the reference system. This increase in thermopower corresponds to the peaks in the transport function at the edge of the conduction band, which is visible in Fig.~\ref{fig:transport_sigma}(b). 
As the temperature increases (600$-$800~K), both maxima shift towards lower concentrations, while increasing the absolute value of thermompower with over $-250$ $\mu$V/K achieved for $\mathrm{Sn_{0.984}Zn_{0.01}Te}$ and $-175$ $\mu$V/K in the case of $\mathrm{Sn_{0.964}Zn_{0.03}Te}$.
In $n$-type transport, we observe a significantly stronger bipolar effect as for $n<10^{20}$ cm$^{-3}$ [see Fig.~\ref{fig:transport_summary_ntpe}(d)] the bipolar onset appears at 400$-$550~K in contrast to $p$-type, where it would start to affect thermopower only above 700 K at similar carrier concentrations [Fig.~\ref{fig:transport_summary_ptype}(d)]. The reason for this difference lies first and foremost in the asymmetry in the transport function, since the $\sigma(E)$ for 1\% and 3\% Zn (Fig. \ref{fig:transport_sigma}) is up to order of magnitude higher in the $p$-type. Secondly, in the case of the lower dopant concentrations, a smaller band curvature is observed for CB than in VB (See Fig. ~\ref{fig:208_bsf_gap_inv}) which additionally introduces an asymmetry of effective masses ($m^*_n>m^*_p$).

Overall, from the two variants mentioned above, $\mathrm{Sn_{0.984}Zn_{0.01}Te}$ yields significantly higher improvement in comparison to I-doped SnTe, as in the case of 3\%, the thermopower enhancement emerges only at very high carrier concentrations [Fig.~\ref{fig:transport_summary_ntpe}(g)]. 
Regarding the sample with the highest Zn content, our calculations indicate that $\mathrm{Sn_{0.914}Zn_{0.08}Te}$ possesses the highest maximum thermopower among the variants studied, while also showing a stable increase in the Seebeck coefficient over the reference system. It's peak thermopower, however, appears only for the low carrier concentration regime ($n<10^{20}$ cm$^{-3}$), reaching over $-300$ $\mu$V/K near 600 K. From the temperature dependence plots we also observe a less notable bipolar effect in comparison to the variants with smaller Zn concentration, due to larger band gap as well as higher conductivity in CB [Fig. \ref{fig:transport_sigma}(b)] which causes smaller asymmetry in $\sigma(E)$ between $n$- and $p$-type.


        \section{Summary and conclusions}

In summary, in this work we presented a theoretical study of the electronic structure and transport properties of Sn$_{0.994-x}$Zn$_x$Te for a wide range of Zn concentrations of  $0.1\%-8\%$ using two complementary methods based on DFT: KKR-CPA and the supercell approach.
Our results show that Zn does not form a resonant level within the valence band, and instead, a possible resonant-like state is present at the edge of the conduction band.
Additionally, analysis of Bloch Spectral Functions and unfolded band structure of the studied supercells prove that Zn does not form an isolated impurity band in this system, however, doping induces a topological phase transition converting Sn$_{0.994-x}$Zn$_x$Te to a conventional insulator between 2\% and 4\% of zinc content. As a consequence, from the TE perspective only the extreme (i.e. either low or very high) dopant concentrations might be favorable considering closing of the band gap near 2\% of Zn.

In spite of the location of an RL within the CB, Zn-doping has extensive influence on the last valence band, which becomes considerably smeared. That decreases the carrier lifetime by up to two orders of magnitude with increasing zinc content. The VB undergoes significant modifications under the influence of dopant, as both band convergence and band flattening occur, with the latter leading to an increase in the effective mass. 
Those results and the position of the RL suggest that the resonant level is not directly responsible for the enhancement of the thermopower in this system, as reported in the literature \cite{Bhat2019Zn:Thermoelectrics}. Calculations of electronic transport performed within the Kubo-Greenwood formalism show that in $p$-type SnTe, zinc reduces the onset of bipolar conduction and causes a general enhancement of the Seebeck coefficient of Zn-doped SnTe in a wide carrier concentration range. The calculated thermopower remains in good agreement compared to the experimental measurements reported in the literature \cite{Bhat2019Zn:Thermoelectrics,Chen2018BandZn-doping}, with the sample $\mathrm{Sn_{0.914}Zn_{0.08}Te}$ achieving the highest Seebeck coefficient of 250 $\mu$V/K for $10^{20}$ cm$^{-3}$ at 800~K.

For the $n$-type material, doping with Zn leads to the formation of a sharp DOS peak at the edge of the conduction band caused by charge redistribution from deeper parts of the CB. Formation of Zn DOS peak which is highly hybridized with the host states and strong distortion of the band structure suggest formation of a resonant level. However, a detailed analysis of the new features caused by Zn doping shows that while zinc states group near the original band of undoped SnTe, high amounts of impurities lead to a complex evolution of the conduction band which does not retain all of the characteristics of a typical resonant level, such as In in SnTe \cite{Wiendlocha2021ResidualSemiconductors}. Instead, new features appear in the band structure near the bottom of the conduction band and the band structure is rebuilt.
Despite that, the calculation of the transport properties of $n$-type SnTe:Zn shows a significant increase in thermopower --  up to 150 $\mu$V/K more \added[]{(Fig.\ref{fig:transport_summary_ntpe})} -- in the Zn-doped variants of SnTe compared to the reference system $\mathrm{SnTe_{0.995}I_{0.005}}$. However, this enhancement is limited to a slightly narrower carrier concentration range in contrast to $p$-type transport. In general, our results show that the thermoelectric performance of SnTe should be greatly improved through Zn doping if $n$-type samples could be synthesized.

\section{Acknowledgements}

This research project was supported by the ''Excellence Initiative – Research University'' program at AGH University of Krakow. The authors gratefully acknowledge the Polish high-performance computing infrastructure PLGrid (HPC Center: ACK Cyfronet AGH) for providing computer facilities and support within computational Grant No. PLG/2025/018404.

\bibliography{main}

\newpage
\onecolumngrid
\section*{Supplemental Material}

\renewcommand{\thefigure}{{S\arabic{figure}}}
\setcounter{figure} 0

\vspace*{24pt}
\noindent
    Supplemental Material contains:\\
    Fig. S1 with comparison of DOS for $\mathrm{Sn_{26}ZnTe_{27}}$ supercell and $\mathrm{Sn_{0.954}Zn_{0.04}Te}$;\\
    Fig. S2 with unfolded band structure for $\mathrm{Sn_{26}ZnTe_{27}}$ supercell showing the effects of structural relaxation and higher energy cutoff;\\
    Fig. S3 with calculated band gap width as a function of Zn content;\\
    Fig.~S4 with calculated band structure with indicated band parity at high symmetry points for $\mathrm{Sn_{27}Te_{27}}$ and $\mathrm{Sn_{26}ZnTe_{27}}$ supercells;\\
    Fig. S5 with calculated DOS and BSF for for SnTe$_{0.995}$I$_{0.005}$.

\vspace*{24pt}

\begin{figure}[hp!]
        \includegraphics[width=0.5\textwidth]{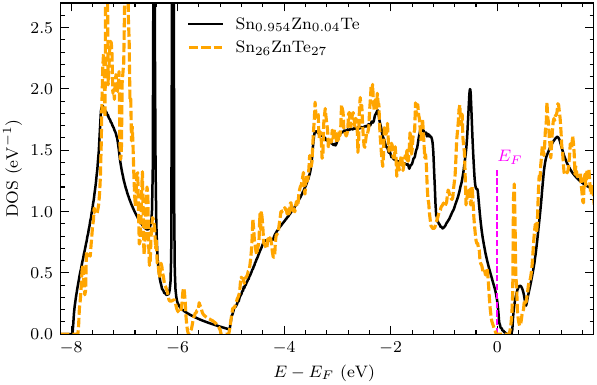}
        \caption{Comparison of the total DOS for $\mathrm{Sn_{0.954}Zn_{0.04}Te}$ and supercell $\mathrm{Sn_{26}ZnTe_{27}}$. }
        \label{fig:001c_dos_comp}
    \end{figure}

    In Fig.~\ref{fig:001c_dos_comp} a comparison of the total density of states (DOS) is presented for $\mathrm{Sn_{0.954}Zn_{0.04}Te}$ and $\mathrm{Sn_{26}ZnTe_{27}}$ supercell (simulating 3.7\% Zn doping) calculated using the KKR-CPA approach and pseudopotential method, respectively. Structure of the supercell was fully optimized, i.e. the relaxation was performed for atomic positions as well as shape of the unit cell. After the relaxation the lattice parameter $a$ increased by 0.993\%. Regarding the atomic positions, position of Zn atom was slightly shifted changing its  distance to the nearest neighbors from 3.16 \AA\ to $2.79-3.28$ \AA. Results from both methods show good agreement only with small differences related to the sharper resonant peak and position of the Fermi level. The latter is caused by lack of Sn vacancies in the supercell which leads to rigid shift of the $E_F$ closer to the band gap. Additionally we also observe small shift of the splitted semicore 3d states from around $-7$ eV in \replaced[comment=R2Q2]{the supercell}{$\mathrm{Sn_{0.954}Zn_{0.04}Te}$} to $-6$ eV in \replaced[comment=R2Q2]{$\mathrm{Sn_{0.954}Zn_{0.04}Te}$}{the supercell}.

    \added[comment=R1Q2]{In case of the supercell calculation using the pseudopotential method tests were performed to verify the influence of the energy cutoff $E_{cutoff}$ for the plane-wave basis set. For this purpose, calculations of electronic structure for the fully relaxed model were performed (without SOC) using $E_{cutoff}=320$ eV and $E_{cutoff}=500$ eV, and the resulting unfolded band structure for those settings is shown in Fig.~\ref{fig:R001} (b) and (c). In general, between both of those cases, no relevant discrepancies are present with the exception of energy range above 1.8 eV, where small differences can be noticed. Considering negligible benefit of the increased $E_{cutoff}$ value, all calculations in this work using the pseudopotential method had the energy cutoff of 320 eV.}

    \added[comment=R1Q3]{To indicate the importance of structural optimization, the unfolded band structure of $\mathrm{Sn_{26}ZnTe_{27}}$ is shown before [Fig.~\ref{fig:R001} (a)] and after [Fig.~\ref{fig:R001} (b)] full relaxation (i.e. atomic positions and the unit cell shape). In the unoptimized structure, the band forming the conduction band minimum crosses the band gap in a small degree in the $\Gamma-$L direction and slightly overlaps with the VB maximum. This behavior is akin to what appears in works by Bhat and Shenoy \cite{Bhat2019Zn:Thermoelectrics,Shenoy2020BiiZT/i} or Zhang \cite{Zhang2021EnhancedDefects}. After structural relaxation the band gap opens Fig.~\ref{fig:R001}(b) and now both VB and CB extrema are located at L point. In addition, inclusion of spin-orbit coupling interaction in calculations primarily affects the band splitting near -2 or 0.5 eV and the region near the band gap is not affected in a meaningful manner.
    }

    Influence of Zn-doping on the band gap's width $E_g$ for Sn$_{0.994-x}$Zn$_{x}$Te is shown in Fig.~\ref{fig:207bg}. Here, the $E_g$ value was estimated by the position of fitted Lorentzian peak. 
    The value of $E_g$ in the TCI phase decreases with increasing Zn concentration from 0\% up to about 2\%, closing the band gap. In this region the band structure near the band edges is inverted.
    Further doping introduces topological phase transition to the trivial phase,  gradually reopening the band gap. For 8\% of dopant the gap returns to its almost original absolute value before the introduction of Zn. 
    As far as the absolute value of the gap is concerned, a non-monotonous behavior of $E_g(x)$ is observed, as in case of Fig.~6 in the main manuscript.
    

    \added[comment=R3Q4]{In Fig.~\ref{fig:401_bandparity} we show the dispersion relation for two supercells $\mathrm{Sn_{27}Te_{27}}$  and $\mathrm{Sn_{26}ZnTe_{27}}$ together with indicated band parity at high symmetry points. Between both structures inversion of the band parity is observed at the L point between the conduction and valence band extrema, which is an additional indicator of a topological phase transition; in this case transition to topologically trivial state via Zn doping.}

    Figure \ref{fig:008_sntei} shows the DOS and two-dimensional projections of Bloch spectral functions for SnTe$_{0.995}$I$_{0.005}$. Compared to the pristine SnTe with 0.6\% of vacancies, this variant differs only in the position of Fermi level, which here was shifted into the conduction band.
    
    \added{    
    \begin{figure}[hp!]
        \includegraphics[width=1\textwidth]{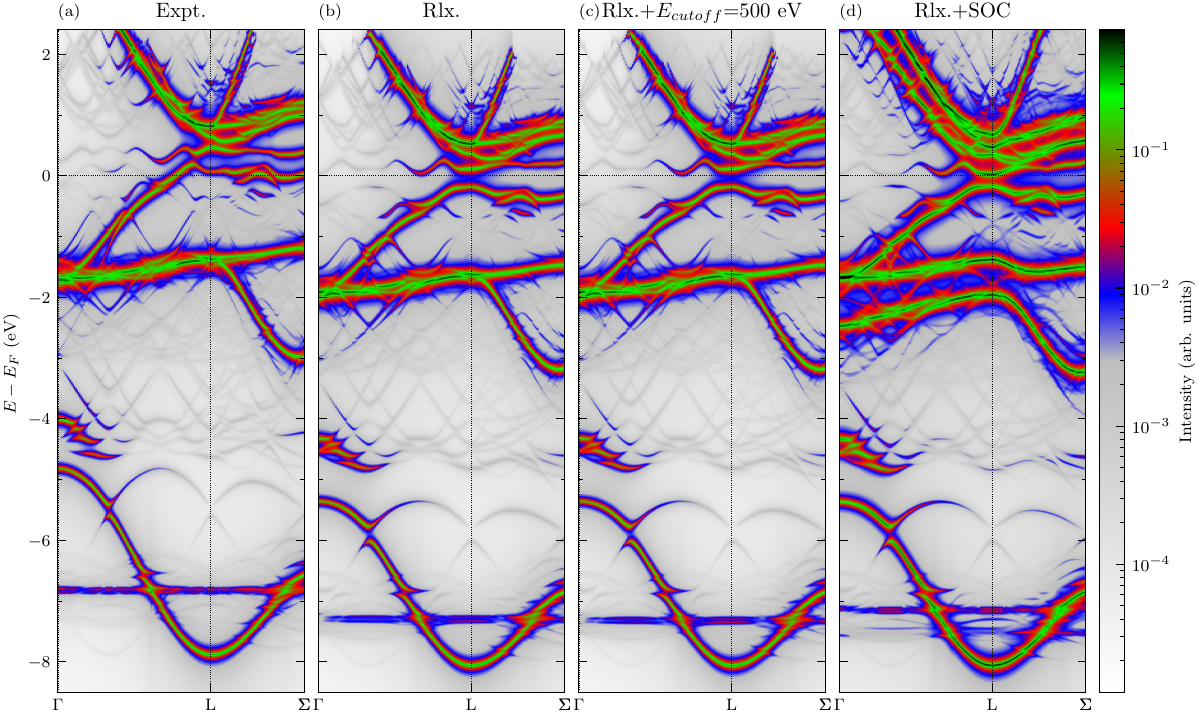}
        \caption{Comparison of unfolded band structure for the $\mathrm{Sn_{26}ZnTe_{27}}$ along $\Gamma$$-$L$-$$\Sigma$ calculated using: (a) experimental crystal structure; (b) fully relaxed crystal structure (unit cell and atomic positions); (c) fully relaxed crystal structure with increased cutoff energy for the plane wave base to 500 eV; (d) fully relaxed crystal structure with spin-orbit interaction included. If not specified otherwise, default energy cutoff of 320 eV was used.}
        \label{fig:R001}
    \end{figure}
    }

    \begin{figure}[hp!]
        \includegraphics[width=0.5\textwidth]{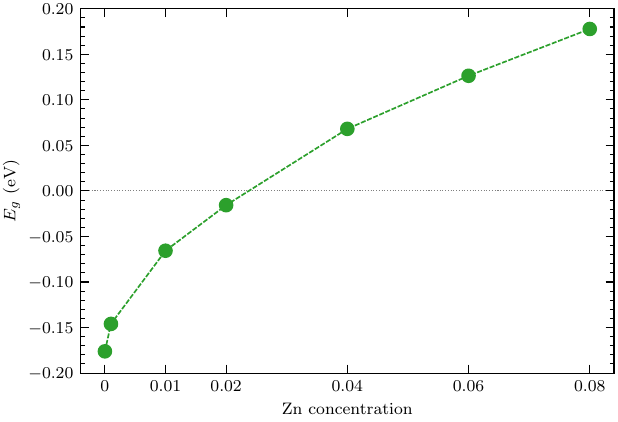}
        \caption{Calculated band gap $E_g(x)$ with respect to the Zn concentration in Sn$_{0.994-x}$Zn$_{x}$Te. The value of $E_g$ in the TCI (inverted) phase is marked as negative. Absolute value of $E_g$ decreases with increasing Zn concentration from 0\% up to about 2\%, closing the band gap. Further doping introduces topological phase transition to the trivial phase,  gradually reopening the band gap. 
        }
        \label{fig:207bg}
    \end{figure}

    \begin{figure*}[hp!]
        \centering
        \includegraphics[width=\textwidth]{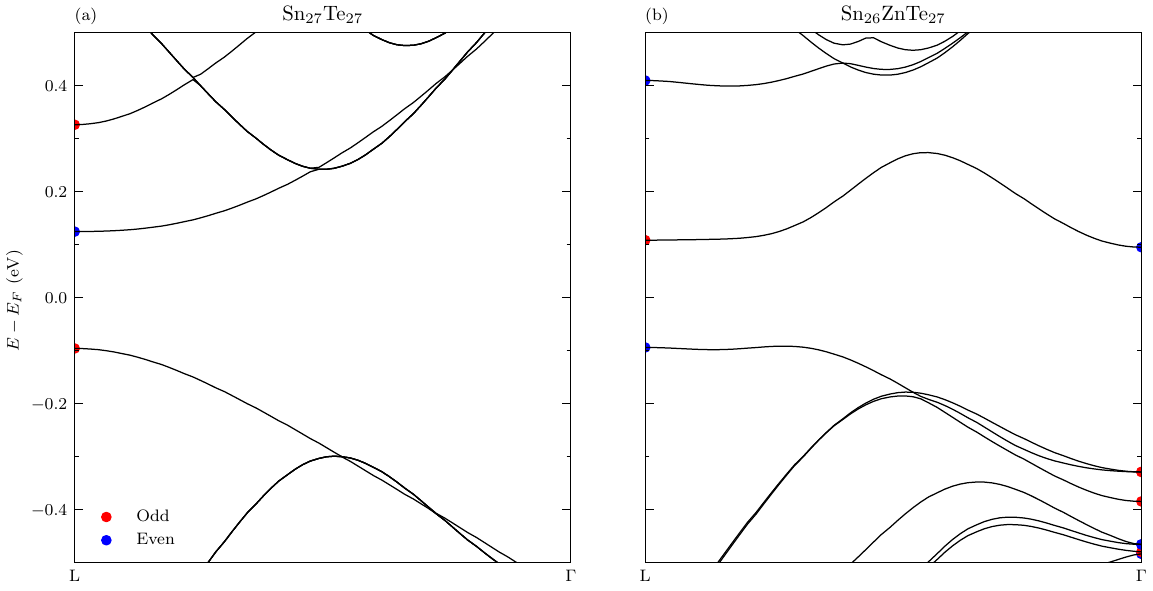}
        \caption{\added{Calculated band structure with indicated band parity at high symmetry points for (a) $\mathrm{Sn_{27}Te_{27}}$ (experimental crystal structure); (b) $\mathrm{Sn_{26}ZnTe_{27}}$ (after structural relaxation). }}    
        \label{fig:401_bandparity}
    \end{figure*}
    
    \begin{figure*}[hp!]
        \centering
        \includegraphics[width=\textwidth]{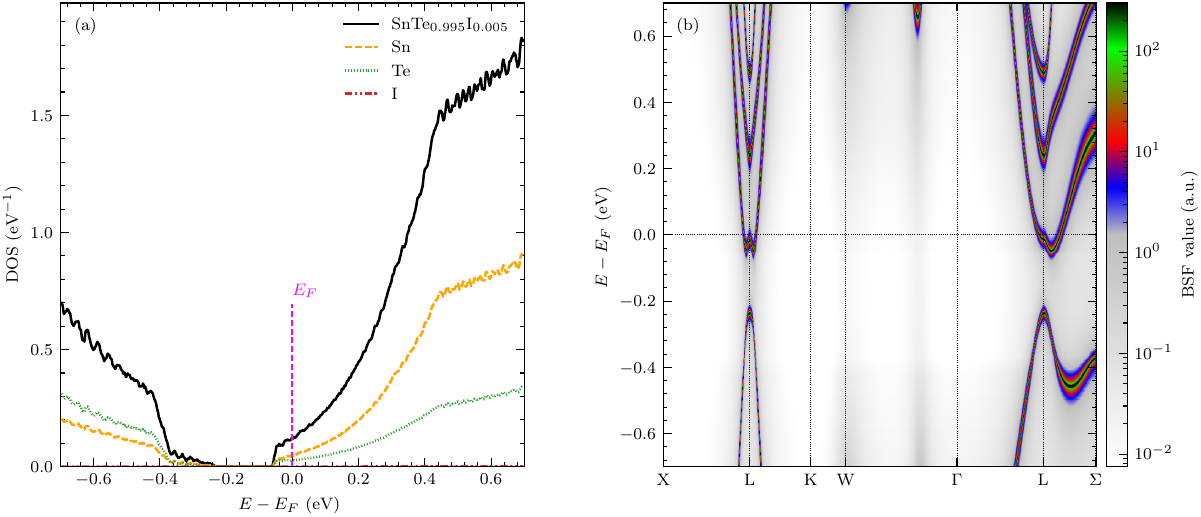}
        \caption{(a) Calculated density of states for SnTe with 0.5\% of I (b) Two-dimensional projection of Bloch spectral functions for SnTe$_{0.995}$I$_{0.005}$. Color representing BSF is in a logarithmic scale, where black color corresponds to values higher than 300 atomic units.}    
        \label{fig:008_sntei}
    \end{figure*}

\end{document}